\documentclass[hyper,letterpaper,notoc]{JHEP3}

\usepackage{epsfig}
\usepackage{amsbsy}
\usepackage{varioref}
\usepackage{pifont}
\usepackage{amsmath}
\usepackage{graphicx}
\usepackage{axodraw}

\def\slashii#1{\setbox0=\hbox{$#1$}             
   \dimen0=\wd0                                 
   \setbox1=\hbox{\sl/} \dimen1=\wd1            
   \ifdim\dimen0>\dimen1                        
      \rlap{\hbox to \dimen0{\hfil\sl/\hfil}}   
      #1                                        
   \else                                        
      \rlap{\hbox to \dimen1{\hfil$#1$\hfil}}   
      \hbox{\sl/}                               
   \fi}                                         %
\def\slashiii#1{\setbox0=\hbox{$#1$}#1\hskip-\wd0\hbox to\wd0{\hss\sl/\/\hss}}
\title{Deconstruction and Elastic $\pi\pi$ Scattering\\
 in Higgsless Models}

\author{R. Sekhar Chivukula and Elizabeth H. Simmons\\
Department of Physics and Astronomy, Michigan State University\\
East Lansing, MI 48824, USA\\
E-mail: sekhar@msu.edu, esimmons@msu.edu}

\author{Hong-Jian He\\
Center for High Energy Physics, Tsinghua University\\
Beijing 100084, China\\
	E-mail: hjhe@mail.tsinghua.edu.cn}

\author{
Masafumi Kurachi\\
C.N. Yang Institute for Theoretical Physics, State University of New York\\
Stony Brook, NY 11794, USA\\
	E-mail: masafumi.kurachi@stonybrook.edu}

\author{
Masaharu Tanabashi\\
Department of Physics, Tohoku University\\
Sendai 980-8578, Japan\\
	E-mail: tanabash@tuhep.phys.tohoku.ac.jp}

\abstract{
We study elastic pion-pion scattering  in global linear moose models and apply the results to a variety of Higgsless models in flat and AdS space using the Equivalence Theorem.   In order to connect the global moose to Higgsless models, we first introduce a block-spin
transformation which corresponds, in the continuum, to the freedom to perform coordinate transformations in the Higgsless model.  We show that it is possible to make an ``f-flat'' deconstruction in which all of the f-constants $f_j$ of the linear moose model are identical; the phenomenologically relevant f-flat models are those in which the coupling constants of the groups at either end of the moose are small -- corresponding to the global linear moose.  In studying pion-pion scattering, we derive various sum rules, including one analogous to the KSRF relation, and use them in evaluating the low-energy and high-energy forms of the leading elastic partial wave scattering amplitudes.  We obtain elastic unitarity bounds as a function of the mass of the lightest $KK$ mode and discuss their physical significance.
\\ \\ 
\centerline{January 9, 2007}}

\keywords{Dimensional Deconstruction, Electroweak Symmetry Breaking, Higgsless Theories}

\preprint{MSUHEP-061206\\
YITP-SB-06-55\\
TU-782}

\begin{document}

\section{Introduction}

Higgsless models \cite{Csaki:2003dt}  break the electroweak symmetry without employing a fundamental scalar  Higgs boson \cite{Higgs:1964ia}. 
Motivated by gauge/gravity duality \cite{Maldacena:1998re,Gubser:1998bc,Witten:1998qj,Aharony:1999ti}, models of this kind may be viewed as ``dual'' to more conventional models of dynamical symmetry breaking 
\cite{Weinberg:1979bn,Susskind:1979ms} such
as ``walking techicolor'' \cite{Holdom:1981rm,Holdom:1985sk,Yamawaki:1986zg,Appelquist:1986an,Appelquist:1987tr,Appelquist:1987fc}.
In these models, the scattering of longitudinally polarized electroweak gauge bosons is unitarized through the exchange of extra vector bosons \cite{SekharChivukula:2001hz,Chivukula:2002ej,Chivukula:2003kq,He:2004zr}, rather than scalars. Based on TeV-scale \cite{Antoniadis:1990ew} compactified $SU(2)^2 \times U(1)$ five-dimensional gauge theories with appropriate boundary conditions \cite{Agashe:2003zs,Csaki:2003zu,Burdman:2003ya,Cacciapaglia:2004jz}, Higgsless models provide effectively unitary descriptions of the electroweak sector beyond the TeV energy scale. They are not, however, renormalizable, and can only be viewed as effective theories valid below a cutoff energy scale inversely proportional to the square of
the five-dimensional gauge-coupling. Above this energy scale, 
some new ``high-energy" completion, which is valid to higher energies, must obtain.

\EPSFIGURE[th]{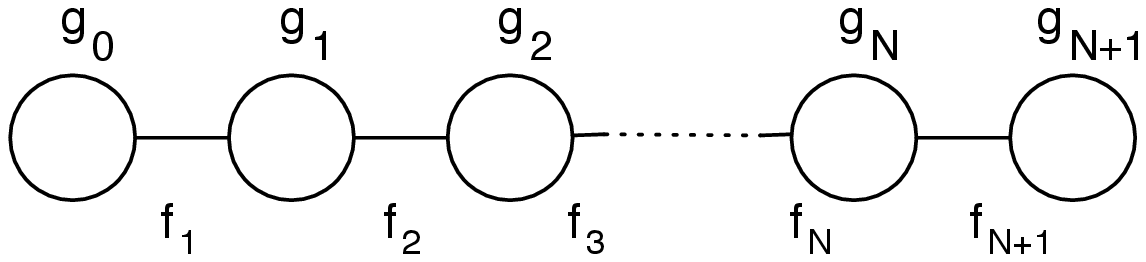,width=0.6\textwidth}
{Deconstructed model, in moose notation \protect\cite{Georgi:1986hf}, corresponding
to an arbitrary five-dimensional gauge theory. The coupling constants ($g_i$)
and $f$-constants ($f_j$) are arbitrary, corresponding, as discussed in the text, 
 to the position-dependent coupling and warp factors chosen.}
 \label{fig:ozone}

Since a compactified five-dimensional
theory gives rise to an infinite tower of massive four-dimensional ``Kaluza-Klein" ($KK$) modes of the 
gauge field, it would seem reasonable to be able to construct a low-energy approximation 
with only a finite number of low-mass fields. Deconstruction 
\cite{Arkani-Hamed:2001ca,Hill:2000mu} provides a realization of this 
expectation and, as illustrated in Fig. \ref{fig:ozone},  
may be interpreted as replacing the continuous fifth dimension with a 
lattice of individual gauge groups (the factors of the semi-simple  
deconstructed four-dimensional gauge group) at separate sites in ``theory space". 
The ``link" variables connecting the gauge groups
at adjacent sites represent symmetry breaking fields which break the groups at adjacent
sites down to the diagonal  subgroup, as conventional
in ``moose notation" \cite{Georgi:1986hf}. In the continuum limit (in which the number of
sites on the deconstructed lattice is taken to infinity), the kinetic energy terms for the link
variables give rise to the
terms in the five-dimensional kinetic energy involving derivatives with respect to
the compactified coordinate.
Deconstructed Higgsless models \cite{Foadi:2003xa,Hirn:2004ze,Casalbuoni:2004id,Chivukula:2004pk,Perelstein:2004sc,Georgi:2004iy,SekharChivukula:2004mu} have been used as tools to compute the general properties of Higgsless theories, and to illustrate the phenomological properties of this class of models.\footnote{Deconstucted Higgsless models with only a few extra vector
bosons  are, formally, equivalent to models of extended electroweak gauge symmetries
\protect\cite{Casalbuoni:1985kq,Casalbuoni:1996qt}  motivated by models of hidden local symmetry \protect\cite{Bando:1985ej,Bando:1985rf,Bando:1987br,Bando:1988ym,Bando:1988br,Harada:2003jx}.}

The $KK$ modes, as represented in the deconstructed model, are obtained
by diagonalizing the gauge field mass-squared matrix resulting from the
symmetry breaking of the deconstructed model. Recalling that this mass matrix
reproduces, in the continuum limit, terms involving derivatives with respect to
the compactified coordinate, we see that the lightest modes will correspond to eigenvectors
which vary the least from site to adjacent site. Therefore, given a particular deconstruction,
it would seem that a related model with fewer sites should suffice to
describe only the lowest modes. This is precisely the reasoning of the Kadanoff-Wilson
``block spin" transformation \cite{Kadanoff:1966wm,Wilson:1971bg} 
applied to compactified five-dimensional gauge theories.

In the first third of this paper, we apply  the block-spin transformation to a deconstructed theory, and
show how this  allows us to obtain
 an alternative deconstruction with fewer factor groups and exhibiting the same 
 low-energy properties. We demonstrate that this freedom to
do block-spin transformations corresponds, in the continuum limit, to the freedom
to describe the continuum theory in different coordinate systems. We apply this technique
to gauge-theories in $AdS_5$, discussing the conventional  ``$g$-flat'' deconstruction \cite{Cheng:2001nh,Abe:2002rj,Falkowski:2002cm,Randall:2002qr} in which all the gauge couplings $g_j$ of the linear moose are identical, a ``conformal" deconstruction corresponding to conformal
coordinates in the continuum,  and an ``$f$-flat" deconstruction in which all of the 
$f$-constants $f_j$ of the linear moose are identical.

\EPSFIGURE[t]{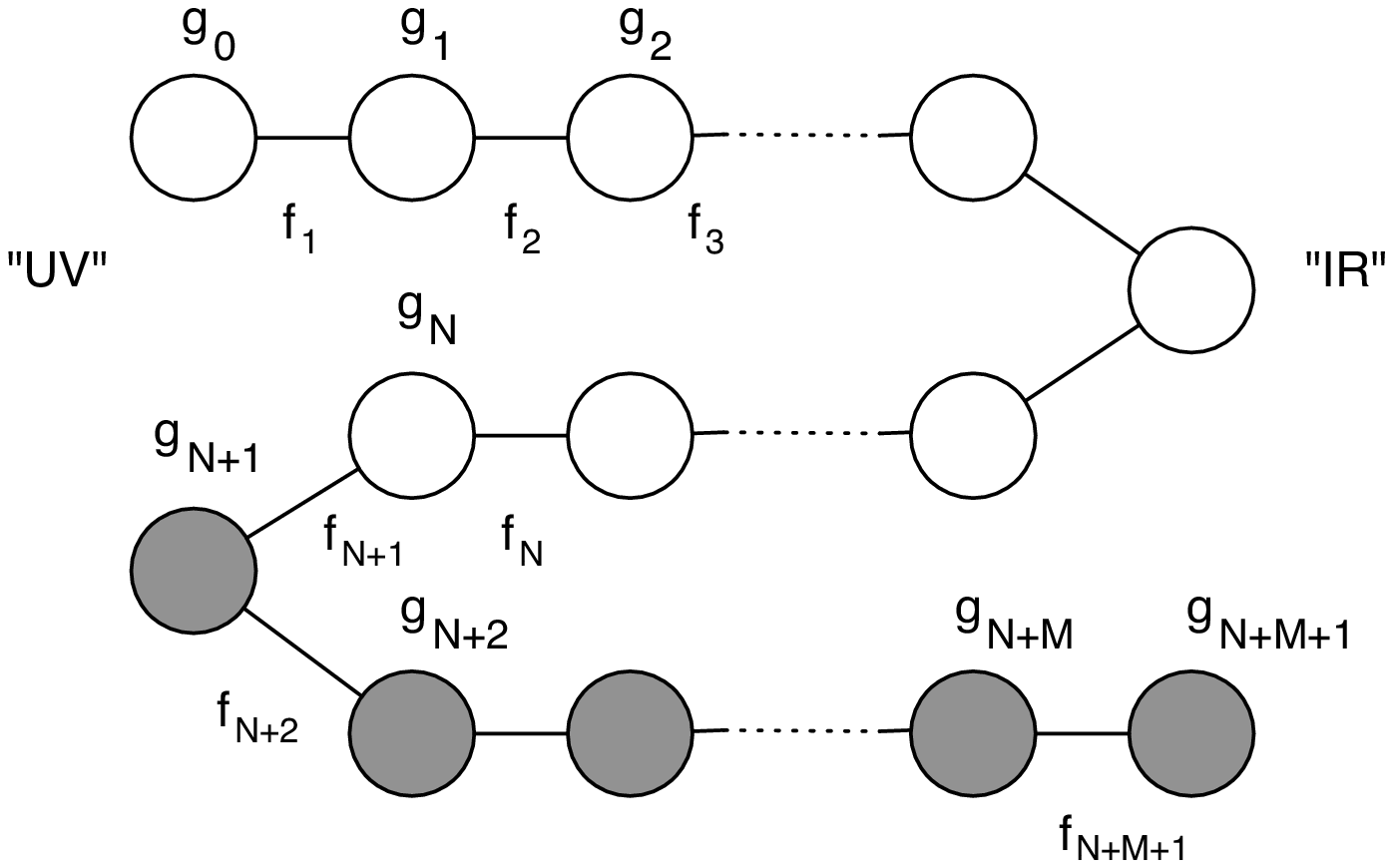,width=0.6\textwidth}
{Moose diagram for $SU(2)_L \times SU(2)_R \times U(1)$ Higgsless models \protect{\cite{Csaki:2003zu,Cacciapaglia:2004jz}}. Open circles represent $SU(2)$ groups, shaded circles $U(1)$ groups.
The light fermions are localized near the ``UV" brane and couple, largely, to the groups at sites 0 and 
$N+1$. Using the $f$-flat deconstruction we find \protect{\cite{Georgi:2004iy,SekharChivukula:2004mu}}
that the phenomenologically relevant models correspond to those with $g_0$ and $g_{N+1}$
small. Using the equivalence theorem \protect{\cite{Cornwall:1974km,Lee:1977eg,Chanowitz:1985hj,Yao:1988aj,Bagger:1990fc,He:1992ng,He:1994yd,He:1997cm}}, the scattering of longitudinally
polarized electroweak gauge bosons corresponds to that of the pions arising from the model
in the limit $g_0,\, g_{N+1} \to 0$ -- the ``global moose" illustrated in Fig. \protect{\ref{fig:oone}}.
\label{fig:SU2SU2U1bulk}}

In the remainder of this paper, we apply these findings to a study of the properties of $W_L W_L$ elastic scattering in $SU(2)^2 \times U(1)$ Higgsless models, which are illustrated in deconstructed form
in Fig. \ref{fig:SU2SU2U1bulk}.  The gauge sector of a phenomenologically-relevant model includes 
 a single massless photon, a light $W^\pm$, a light $Z$, and additional $KK$ modes 
 with masses substantially higher than those of the $W$ and $Z$. Using an appropriate
 block-spin transformation, we may analyze any model in an $f$-flat deconstruction and,
 as shown in refs. \cite{Georgi:2004iy,SekharChivukula:2004mu}, 
the phenomenologically relevant models correspond to those with $g_0$ and $g_{N+1}$
small.\footnote{In addition, the fermions 
 must be slightly delocalized \protect{\cite{Cacciapaglia:2004rb,Cacciapaglia:2005pa,Foadi:2004ps,Foadi:2005hz,SekharChivukula:2005xm}} away from the ``UV" brane in order to satisfy
 both the constraints of precision electroweak observables
 and those of unitarity in $W_L W_L$ scattering.  This constraint will not be relevant to the
 issue of $W_L W_L$ elastic scattering, and will therefore play no role in our analysis.}

\EPSFIGURE[th]{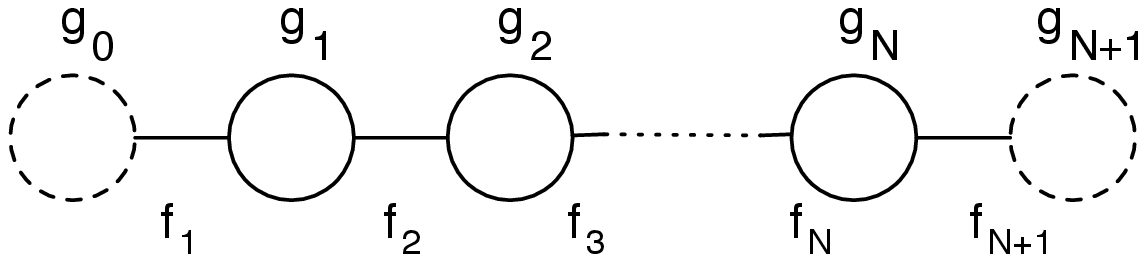,width=0.6\textwidth}
{The global linear moose for which elastic $\pi\pi$ scattering is 
studied in this paper.  Sites $0$ and
$N+1$ are global groups ($g_0=g_{N+1}=0$) and sites $1$ to $N$ are $SU(2)$ gauge groups.
One obtains the deconstructed form of either flat space or AdS models depending on the couplings
and $f$-constants chosen. For simplicity, we will only consider ``$f$-flat'' mooses, for which
$f = v\,\sqrt{N+1}$, in detail -- the generalization to nonflat $f$'s is straightforward, using the block spin transformations and f-flat deconstruction of Section 2.
\label{fig:oone}}

According to the equivalence theorem\footnote{The generalization
of the equivalence theorem to compactified five-dimensional theories is discussed
in \protect{\cite{SekharChivukula:2001hz}}.} \cite{Cornwall:1974km,Lee:1977eg,Chanowitz:1985hj,Yao:1988aj,Bagger:1990fc,He:1992ng,He:1994yd,He:1993qa,He:1997cm}, at energies $E \gg M_W$ 
the scattering of longitudinally
polarized electroweak gauge bosons corresponds to that of the pions arising from the model
in the limit $g_0,\, g_{N+1} \to 0$ -- the ``global moose" illustrated in Fig. \protect{\ref{fig:oone}.
Since the global linear moose has an overall $SU(2)_L
\times SU(2)_R$ global symmetry which is spontaneously broken, this theory will have 3 exact
Goldstone bosons. By examining the ``dual moose" (formed by, formally, exchanging the values of the gauge-couplings and the $f$-constants  \cite{Sfetsos:2001qb,SekharChivukula:2004mu}), we immediately see that the massless
pion field $\pi(x)$ may be written
\begin{equation}
\pi(x) = v\, \sum_{m=1}^{N+1} \frac{\pi_m(x)}{f_m}~,
\label{eq:pionwv}
\end{equation}
where the $\pi_i(x)$ are the pions in the $i$th link of the moose and $v$
is the $F$-constant of Goldstone boson field $\pi(x)$
\begin{equation}
\frac{1}{v^2} = \sum_{m=1}^{N+1} \frac{1}{f^2_m}~.
\end{equation}

We turn, in section 3, to deriving the form of the amplitudes for elastic pion-pion scattering in global linear moose models.  
In section 4, we find the leading partial-wave amplitudes, examine their limiting forms in the case of scattering energies well below or well above the mass of the lightest KK modes and confirm 
an alternative formulation of the low-energy scattering amplitudes in terms of the Longhitano parameters.  We derive various sum rules,  including one \cite{DaRold:2005zs} 
analogous to the KSRF relation of QCD \cite{Kawarabayashi:1966kd,Riazuddin:1966sw}, and
use them to evaluate the leading partial wave amplitudes  in several continuum Higgsless models in flat and warped space.  

Viewed in terms of the AdS/CFT conjecture \cite{Maldacena:1998re,Gubser:1998bc,Witten:1998qj,Aharony:1999ti}, the models described here may be viewed as approximations of
5-D dual to a theory with chiral symmetry breaking dynamics. From this perspective,  
our results correspond to the chiral dynamics investigated previously in AdS/QCD 
\cite{Babington:2003vm,Evans:2004ia,Erlich:2005qh,Sakai:2004cn,Hirn:2005nr,Sakai:2005yt}. 
In terms of more conventional low-energy theories of 
QCD, the models described here generalize those of $\pi \pi$
scattering which incorporate vector mesons \cite{Harada:1995dc}.

Finally, in section 6,  we study the elastic unitarity limits on the range of validity of the linear moose model and conclude that this provides a useful guide only for models in which the lightest KK mode has a mass greater than about 700 GeV.  Otherwise, a coupled-channel analysis including two-body inelastic modes must be performed.  Section 7 summarizes our conclusions.


\section{From the Block-spin transformation to an  F-flat deconstruction}
\label{sec:thre}

\subsection{The Continuum Limit of An Arbitrary Moose}
\label{sec:two}

We begin by considering the continuum limit of a general
linear moose model of the form shown in Fig. \ref{fig:ozone}. The action
for the moose model at $O(p^2)$  is given by
\begin{equation}
  {\cal S} =
   -\int d^4x\,\sum_{j=0}^{N+1} \dfrac{1}{2g_j^2} \mbox{tr}\left(
    F^j_{\mu\nu} F^{j\mu\nu}
    \right)+
   \int d^4x\,\sum_{j=1}^{N+1} \frac{f^2_j}{4} \mbox{tr}\left(
    (D_\mu U_j)^\dagger (D^\mu U_j) \right)~,
\label{eq:action}
\end{equation}
with
\begin{equation}
  D_\mu U_j = \partial_\mu U_j - i A^{j-1}_\mu U_j 
                               + i U_j A^{j}_\mu,
\end{equation}
and where, for the moment,  all  gauge fields $A^j_\mu$ $(j=0,1,2,\cdots, N+1)$ are dynamical
with the coupling constants $g_j$. 
The gauge groups at the various sites are the same and, for simplicity,\footnote{The
generalization to an arbitrary gauge group is straightforward.}  we
will assume that they are all $SU(n)$ for some $n$. The link fields $U_j$ are non-linear
sigma model fields appropriate to the coset space $SU(n)_{j-1} \times SU(n)_j / SU(n)_{\{j-1\}+\{j\} }$, which
break each set of adjacent groups to their diagonal subgroup, and have the corresponding
decay-constants $f_j$.

To take the continuum limit, we relabel the couplings and decay-constants
by\footnote{Implicitly in Fig. \protect\ref{fig:ozone} we have assumed Neumann boundary
conditions for the continuum limit gauge-theory at the endpoints of the interval
in the fifth dimension. Dirichlet boundary conditions may be imposed by taking
$g_0$ or $g_{N+1}$ to be zero, in which case the corresponding term(s) are excluded
in the sum in Eq. (\protect\ref{eq:hi}). Alternatively, brane-localized gauge kinetic energy
terms may be obtained by holding the corresponding coupling fixed, and not scaling
according to Eq. (\protect\ref{eq:hi}) -- such couplings are excluded from the block-spin
transformations.}
\begin{eqnarray}
g_i = g\,\sqrt{N+2}\, \kappa_i\ \ \ \ {1\over g^2} = \sum_{i=0}^{N+1}{1\over g^2_i}~, 
\label{eq:hi}\\
f_i = f\, \sqrt{N+1}\, h_i\ \ \ \ {1\over f^2} = \sum_{i=1}^{N+1}{1\over f^2_i}~.
\label{eq:kappi}
\end{eqnarray}
The continuum limit corresponds to holding $f$ and $g$ fixed, and $\kappa_i$
and $h_i$ finite, while taking the $N\to \infty$. Note that Eqs. (\ref{eq:hi}) and
(\ref{eq:kappi})
imply that
\begin{equation}
{1\over N+2}\,\sum_{i=0}^{N+1}{1\over \kappa^2_i}={1\over N+1}\,\sum_{i=1}^{N+1}{1\over h^2_i}=1~.
\end{equation}

Defining the dimensionless coordinate  $y_i \equiv  {i\over N+1}~,$
we find the continuum relations:
\begin{eqnarray}
y_i &\to& y~,\qquad
\Delta y = {1\over N+1} \to dy~,\qquad
{1\over N+1} \sum_{i=0}^{N+1} \to \int^1_0 dy~,\\
A^j_\mu (x) &\to& A_\mu(x,y)~,\qquad
(N+1) \left[A^{j+1}_\mu(x)-A^j_\mu(x)\right] \to{\partial A_\mu \over \partial y}~\\
\kappa_i &\to& \kappa(y) ~,\qquad 
{1\over N+2}\,\sum_{i=0}^{N+1}{1\over \kappa^2_i} \to \int^1_0 dy\,{1\over \kappa^2(y)} =1~,
\label{eq:kappaeq}\\
h_i &\to& h(y)~,\qquad
{1\over N+1}\,\sum_{i=1}^{N+1}{1\over h^2_i} \to \int^1_0 dy\,{1\over h^2(y)} =1~.
\label{eq:heq}
\end{eqnarray}

In unitary gauge ($U_j \equiv {\cal I}$), in the continuum limit, the action of Eq. (\ref{eq:action}) 
becomes
\begin{equation}
{\cal S}_5 = \int d^4x\, dy \left[
-\,{1\over 2\, g^2 \kappa^2(y)}\, \mbox{tr}(F_{\mu\nu} F^{\mu\nu})
+{f^2 h^2(y) \over 4} \mbox{tr}\,(F_{\mu y} F^\mu{}_{y})\right]~,
\label{eq:5daction}
\end{equation}
where we have used the appropriate notation for a five-dimensional
gauge theory in unitary gauge ($A_y (x,y) \equiv 0$) in which
\begin{equation}
F_{\mu y} = -\partial_y A_\mu~, \ \ \ \ \ F^\mu{}_y = -\partial_y A^\mu~.
\end{equation}
The form of the action in Eq. (\ref{eq:5daction}) is equivalent to
the form given in \cite{Sfetsos:2001qb,He:2004zr}, and may be
interpreted as arising from a position-dependent five-dimensional gauge-coupling
$g^2_5(y) = g^2 \kappa^2(y)$ in a space with a metric\footnote{Note that, since we
are using a dimensionless coordinate $y$, the interval length $ds$ is dimensionless
as well. From the form of the metric, we see that $gf$ provides the dimensional
factor relating dimensionless and conventional distance.}  given by
\begin{equation}
ds^2 = {(gf)^2 \over 4} \kappa^2(y) h^2(y) \eta_{\mu \nu} dx^\mu dx^\nu - dy^2~.
\end{equation}

\subsection{The Block-Spin Transformation \& The Continuum Limit}

\EPSFIGURE[th]{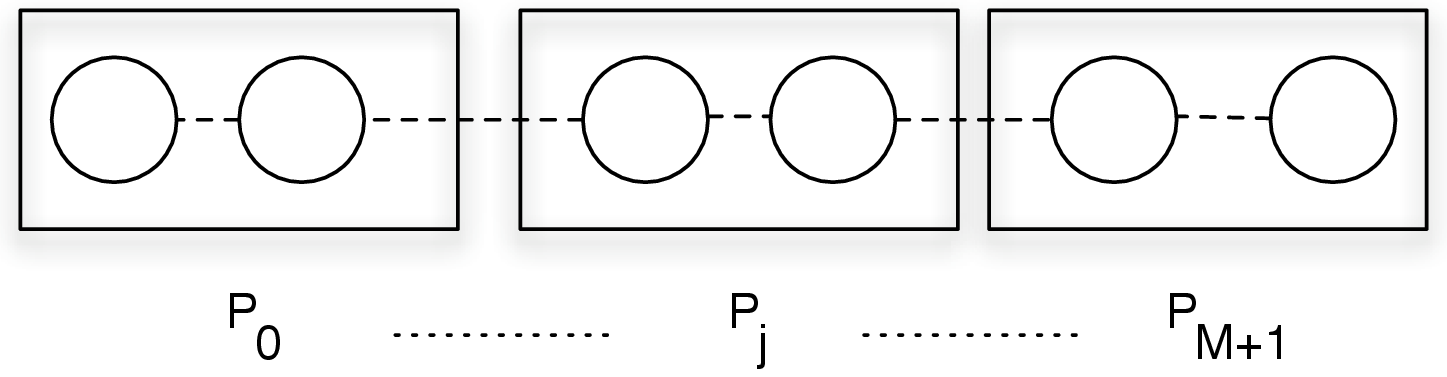,width=0.8\textwidth}
{A block-spin transformation is defined
by a partitioning of the links and sites of the moose shown
in Fig. \protect\ref{fig:ozone} . We will label the partitions -- the set of
links and sites -- by $P_j$, $j=0,1,\ldots,M+1$. We will label the
number of sites in each partition ${\cal P}_j$.\label{fig:two}}

Now let us consider an arbitrary partition of the $N$-site linear moose as shown in
Fig. \ref{fig:two}. As discussed in the introduction,
we expect that the low-lying modes of the original linear moose model --
those whose amplitude varies little within each partition -- 
should be reproduced by an $M$-site model ($M < N$) with appropriately
defined couplings and $f$-constants. By considering the coupling for
the unbroken ``diagonal" subgroup
of any given partition, and the corresponding construction for the ``dual moose"
\cite{Sfetsos:2001qb,SekharChivukula:2004mu} formally defined by exchanging
couplings and decay constants, we  define the couplings\footnote{Both of the relations
in Eq. (\protect{\ref{eq:blockspin}}) have a simple interpretation in terms of a
system of masses and springs whose eigenfrequencies correspond to the
masses of the gauge-bosons in moose models \protect{\cite{Georgi:2004iy}}.}
for the block-spin moose corresponding to the partition of Fig. \ref{fig:two} by
\begin{equation}
{1\over \tilde{g}^2_j} = \sum_{i\in P_j} {1\over g^2_i}~,\ \ \ \ \ \ 
{1\over \tilde{f}^2_j} = \sum_{i\in P_j} {1\over f^2_i}~.
\label{eq:blockspin}
\end{equation}
Note the consistency relation
\begin{equation}
\sum_{j=0}^{M+1} {\cal P}_j = N+2~,
\label{eq:N}
\end{equation}
where ${\cal P}_j$ denotes the number of sites in the partition
$P_j$. 

To clarify the relation between the original and block-spin moose, 
we consider the simultaneous continuum limit of both, which
corresponds to $N\gg M \to \infty$.  In this limit, formally, all $KK$
modes are ``low-lying" and we expect the same continuum limit
for the original and the block-spin moose. Defining the continuum coordinate
\begin{equation}
z_j = {j \over M+1} \to z~.
\end{equation}
The consistency relation (\ref{eq:N}) may be written
\begin{equation}
\sum_{j=0}^{M+1} {1\over M+1}\,\left[{M+1 \over N+2}\,{\cal P}_j \right]=1~,
\end{equation}
and we therefore expect that the quantity
\begin{equation}
\ell_j = {M+1\over N+2}\,{\cal P}_j   \to \ell(z)~,
\end{equation}
remains finite in the limit $N \gg M \to \infty$
and satisfies the relation
\begin{equation}
\int_0^1 \ell(z) dz = 1~.
\label{eq:ellnorm}
\end{equation}

The $y$ coordinate corresponding to the right-hand end of partition $P_j$ is given
by
\begin{equation}
y_j = {1\over N+1} \left(\sum_{l=0}^j {\cal P}_l -1\right) \approx 
{1\over M+1}\,\sum_{l=0}^j \ell_l~,
\end{equation}
where the second equality holds to order $1/N$ or $1/M$.
In the $N \gg M \to \infty$ limit, therefore, we find
\begin{equation}
y=\int^z_0 dz'\,\ell(z')~,
\end{equation}
or, equivalently,
\begin{equation}
{dy\over dz} = \ell(z)~.
\label{eq:yzeq}
\end{equation}
This shows how the continuum coordinates of the original and block-spinned
models are related.

\subsection{Consistency of The Block-Spin Transformation}

We may now compute the couplings and $f$-constants in the continuum
limit of the block-spin moose. As $N \gg M \to \infty$, we will assume that the couplings
in the original model are sufficiently smooth, and $M$ is sufficiently large, that
all of the couplings or $f$-constants in a {\it given} partition $P_i$ are approximately
equal. Defining 
\begin{equation}
\tilde{g}_j = g \sqrt{M+2} \,\tilde{\kappa}_j~,\ \ \ \ \ \ 
\tilde{f}_j = f \sqrt{M+1}\, \tilde{h}_j~,
\end{equation}
Eqs. (\ref{eq:hi}), (\ref{eq:kappi}), and (\ref{eq:blockspin}) then imply (to order $1/M$) that
\begin{equation}
{1\over \tilde{\kappa}^2_j} = {(M+1){\cal P}_j \over (N+2) \kappa^2_{i\in P_j}}
= {\ell_j \over \kappa^2_{i\in P_j}}~.
\end{equation}
In the continuum limit $N \gg M \to \infty$, therefore, we find that
$\tilde{\kappa}_j \to \tilde{\kappa}(z)$ with
\begin{equation}
{1\over \tilde{\kappa}^2} = {\ell \over \kappa^2}~.
\label{eq:kappablock}
\end{equation}
Similarly, we find $\tilde{h}_j \to \tilde{h}(z)$ and
\begin{equation}
\tilde{h}^2 = {h^2 \over \ell}~.
\label{eq:hblock}
\end{equation}

Eqns. (\ref{eq:kappablock}) and (\ref{eq:hblock}) imply that the action of 
of Eq. (\ref{eq:5daction}) becomes 
\begin{equation}
{\cal S}_5 \to 
\int d^4x\, dz \left[
-\,{\ell(z)\over 2\, g^2 \kappa^2(z)}\, \mbox{tr}(F_{\mu\nu} F^{\mu\nu})
+{f^2 h^2(z) \over 4\,\ell(z)} \mbox{tr}\,(F_{\mu z} F^\mu_{z})\right]~.
\label{eq:5dactionblock}
\end{equation}
under an arbitrary block-spin transformation.  Moreover, the block-spin transformation
is clearly equivalent to a change of coordinates described by Eq. 
 (\ref{eq:yzeq}) along with the relationship
\begin{equation}
F_{\mu z} = \ell\, F_{\mu y}~,
\end{equation}
which follows from the same transformation.
Note that the block-spin transformations of Eqs. (\ref{eq:kappablock}) 
and (\ref{eq:hblock}) are such that
\begin{equation}
{\tilde{\kappa}\over \tilde{h}}={\kappa \over h}~.
\end{equation}
%

\subsection{Applying the Block-Spin Transformation}

\subsubsection{Flat-Space Gauge Theories}

Before  considering the block-spin transformation
associated with gauge theories in $AdS_5$, it is interesting to
check that the block-spin transformation produces reasonable results in flat space. A gauge theory
with constant coupling in flat space can be deconstructed 
\cite{Arkani-Hamed:2001ca,Hill:2000mu} to a linear
moose with constant couplings and $f$-constants: $\kappa =h \equiv
1$. Consider taking such a moose with $N+2$ sites and partitioning the sites
into $M+2$ sets each containing $a$ links, {\it i.e.}
\begin{equation}
(N+2) = a\,(M+2)~, \ \ \ \ \ {\cal P}_j \equiv a~.
\end{equation}
In this case, in the $N\gg M \to \infty$ limit we find
\begin{equation}
\ell_j = {M+1 \over N+2} {\cal P}_j \to \ell(y)\equiv 1
\end{equation}
and therefore $\tilde{\kappa}=\kappa$ and $\tilde{h} = h$; in other words, 
we reproduce the form of the deconstructed moose that we started with.
As shown in \cite{Arkani-Hamed:2001ca,Hill:2000mu}, both the original and block-spinned mooses
reproduce the properties of the continuum 5d gauge theory (and therefore
of one another) so long as one considers only the
properties of low-lying $KK$ levels such that $n_{KK} \ll M,\,N$.

\subsubsection{Gauge Theories in $AdS_5$: the traditional ``$g$-flat" construction}

A gauge theory in $AdS_5$ is usually deconstructed to a linear moose that is chosen to have 
constant gauge-couplings and site-dependent VEVs
\cite{Cheng:2001nh,Abe:2002rj,Falkowski:2002cm,Randall:2002qr}. We will refer to this as a ``$g$-flat" deconstruction.  In particular,
one deconstructs the gauge theory such that
\begin{equation}
g^2_i = (N+2) g^2~,\ \ \ \ \ f^2_i = (N+1) f^2 \left({e^b-1 \over b}\right) e^{-b y_i}~.
\label{eq:adsi}
\end{equation}
In this equation,  $g$ is the coupling constant of the unbroken low-energy
4-d gauge theory which exists if (as implicit in the moose shown
in Fig. \ref{fig:ozone}) one chooses Neumann boundary conditions
for the gauge fields at $y=0,\,1$.  Likewise, $f$ is a fixed low-energy scale corresponding to the
$f$-constant of the Goldstone boson which would exist if one chose Dirichlet
boundary conditions for the gauge fields at $y=0,\,1$).   The constant  $b$ is 
related to the $AdS_5$ geometry chosen. From Eqs. (\ref{eq:adsi}) and (\ref{eq:hi}, \ref{eq:kappi}) we see that this $g$-flat deconstruction corresponds to 
\begin{equation}
\kappa^2(y) \equiv 1~,\ \ \ \ \ h^2(y) = {1-e^{-b} \over b}\,e^{b(1-y)}~.
\label{eq:hkappaoriginal}
\end{equation}

The constants of the deconstructed theory, $g$, $b$, and $f$, can be
related to the characteristics of the continuum theory as follows. Take the $AdS_5$ metric
to be
\begin{equation}
ds^2 = e^{-2k r_c \pi y} \eta_{\mu\nu} dx^\mu dx^\nu - r^2_c \pi^2 dy^2~,
\label{eq:adsmetrici}
\end{equation}
where $k$ is the $AdS$ curvature, $r_c$ is related to the size of the interval
in the fifth dimension, and the dimensionless coordinate $y$ runs from
0 to 1. The action for a 5d gauge theory with coupling $g_5$ in this
background is
\begin{eqnarray}
{\cal S}_5 &=& -\,{1\over 2 g^2_5} \int d^4 x\,dy\, \sqrt{|G|}\, {\mbox tr}(F_{MN} F^{MN})~,\\
&=& -{1\over 2} \int d^4x\,dy\,{r_c \pi \over g^2_5}\,{\mbox tr}(F_{\mu\nu} F^{\mu\nu})
+ \int d^4x\,dy\, {e^{-2kr_c \pi y}\over g^2_5 r_c \pi}\,{\mbox tr}(F_{\mu y} F^{\mu}_{y})~,
\label{eq:adsactioni}
\end{eqnarray}
where $G_{MN}$ is the metric of Eq. (\ref{eq:adsmetrici}). Comparing the actions
of Eqs. (\ref{eq:5daction}) and (\ref{eq:adsactioni}), and keeping in mind
Eq. (\ref{eq:hkappaoriginal}), we find the relations
\begin{equation}
g^2  =  {g^2_5 \over r_c \pi}~,\qquad 
b =  2 k r_c \pi~, \qquad
f^2  =  {8k \over g^2_5}\, {1\over e^{2kr_c \pi} -1}~.
\end{equation}
For large $b$, these expressions may be inverted to yield
\begin{equation}
k  =  {g f \sqrt{b} \over 4}\,e^{b/2}~,\qquad
g^2_5  =   {2 g \sqrt{b} \over f}\, e^{-b/2}~,\qquad
{1\over r_c \pi}  =   {g f \over 2 \sqrt{b}}\,e^{b/2}~.
\end{equation}

\subsubsection{Conformal Block-Spin}

Alternatively, in the continuum one may use conformal coordinates
to define the $AdS_5$ geometry. Here one takes the metric
to be
\begin{equation}
ds^2 = \left({R \over Z}\right)^2(\eta_{\mu\nu} dx^\mu dx^\nu - dZ^2)~,
\label{eq:conformalmetric}
\end{equation}
where the conformal coordinate may be taken to run from $R < Z < R'$. The
action for a gauge theory becomes
\begin{equation}
{\cal S}_5 = -\,{1\over 2 g^2_5} \int d^4x\, dZ\,\left({R \over Z}\right)
\left[ {\mbox tr}(F_{\mu\nu} F^{\mu\nu}) - 2\, {\mbox tr}(F_{\mu Z} F^{\mu}_{Z})\right]~.
\label{eq:adsconformal}
\end{equation}

In order to perform a block-spin transformation which yields this form
of the action, we require that the coefficients of the 4-d gauge kinetic energy
and ``link" terms be proportional
\begin{equation}
{1\over \tilde{\kappa}^2} = {\ell \over \kappa^2}\, \propto\, 
\tilde{h}^2 ={h^2 \over \ell}~.
\label{eq:conformalcond}
\end{equation}
Hence, we take
\begin{equation}
\ell = A\, \kappa\, h~,
\end{equation}
where $A$ is determined by the normalization condition Eq. (\ref{eq:ellnorm}).
Integrating Eq. (\ref{eq:yzeq}) and normalizing the result, we find
\begin{equation}
\ell(z)= {2\over b}(e^{b/2}-1)e^{-by/2}~,
\end{equation}
or
\begin{equation}
z={e^{by/2}-1 \over e^{b/2}-1}~.
\end{equation}
Calculating $\tilde{\kappa}$ and $\tilde{h}$, we then find
\begin{eqnarray}
{1\over g^2 \tilde{\kappa}^2(z)} & = & \left({2\over g^2 b}\right)\, {(1-e^{-b/2}) \over
e^{-b/2}+(1-e^{-b/2})z}~,\\
{f^2\tilde{h}^2(z)} & = & \left({f^2\over 2}\right) \, {1+e^{-b/2} \over 
e^{-b/2}+(1-e^{-b/2})z~}~.
\end{eqnarray}

Comparing the actions of Eqs. (\ref{eq:5daction}) and (\ref{eq:adsconformal}), we find
that the two are equivalent under the identifications
\begin{eqnarray}
Z & = & R' \left( e^{-b/2}+ (1-e^{-b/2})z\right)~,\\
g^2 &=& {g^2_5 \over R \log(R'/R)}~,\\
b & = & 2 \log(R'/R)~,\\
f^2 & = & {8 R \over g^2_5 ({R'}^2-R^2)}~.
\end{eqnarray}
For large $b$ ($R' \gg R$), these may be inverted to yield
\begin{equation}
R' = R \,e^{b/2} = {4 \over gf\sqrt{b}}~.
\end{equation}
The coordinates $y$ and $Z$ are related by
\begin{equation}
Z={1\over k} e^{k r_c \pi y}~,
\end{equation}
and the parameters $(R,R')$ are related to $(k, r_c \pi)$ through
\begin{eqnarray}
R &=& {1\over k}~,\\
r_c \pi & = & R \log(R'/R)~.
\end{eqnarray}

\subsubsection{$f$-Flat Deconstruction}

Alternatively, as relevant to our discussion of elastic $\pi\pi$ scattering, we can choose to block-spin the warped-space theory such that $\tilde{h}$ is
constant -- {\it i.e.} we can make an $f$-flat deconstruction.  Since $\tilde{h}^2 = h^2/\ell$, we take
\begin{equation}
{dy\over dz}= \ell = A\, h^2 =  A\, \left({e^b - 1 \over b}\right) e^{-by}~,
\label{eq:ffflat}
\end{equation}
where the constant is determined by the normalization condition (\ref{eq:ellnorm}). As
a result of Eq. (\ref{eq:heq}) we find $A=1$ and, therefore $\tilde{h}(z)=1$.
Integrating Eq. (\ref{eq:ffflat})  we find
\begin{equation}
z={e^{by} -1\over e^b-1}~,
\end{equation}
and, therefore,
\begin{equation}
\tilde{\kappa}^2={1\over \ell} = {b\left[e^{-b}+(1-e^{-b})z\right]\over 1-e^{-b}}~.
\end{equation}
The $f$-flat deconstruction is particularly interesting in the
case of Dirichlet boundary conditions ($g_0=g_{N+1}=0$). In this
case the pions that result are the zero-modes of the ``dual moose"
\cite{Sfetsos:2001qb,SekharChivukula:2004mu}, as in Eq. (\ref{eq:pionwv}), and their
wavefunctions are position independent.

Having demonstrated that any warped-space 5-dimensional model may be deconstructed to a linear moose with all the $f_m$ equal, it is now clear that  our analysis of $\pi\pi$ elastic scattering for the 
$f$-flat 
global linear moose will apply, in the continuum limit, to models with arbitrary background 5-D geometry,
spatially dependent gauge-couplings, and brane kinetic energy terms for the gauge-bosons.


\section{Elastic Pion-Pion Scattering Amplitude in the Global Linear Moose}

\EPSFIGURE[th]{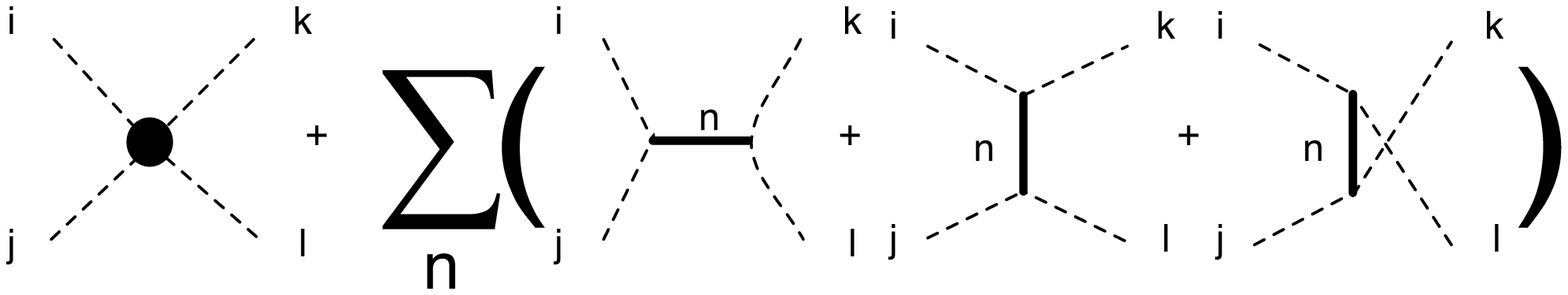,width=0.8\textwidth}
{Tree-level contributions to the scattering amplitude $\pi^i \pi^j \to \pi^k \pi^l$
in the global linear moose. The first term arises from the four-pion contact interactions
from each of the link nonlinear sigma-model kinetic energy terms, and the last three represent
$s$, $t$, and $u$-channel exchange of the massive vector boson fields.
\label{fig:ttwo}
}

Having established the context of our analysis of elastic pion-pion scattering, we
now proceed to calculate the tree-level contributions to the scattering amplitude 
$\pi^i \pi^j \to \pi^k \pi^l$ in the global linear moose illustrated in Figure 3.  Using isospin invariance, 
the form of the pion scattering amplitude may be written
\begin{equation}
i{\cal A}(\pi^i \pi^j \to \pi^k \pi^l)  = A(s,t,u)\delta^{ij} \delta^{kl} + A(t,s,u) \delta^{ik} \delta^{jl}
+ A(u,t,s)\delta^{il} \delta^{jk}~.
\label{eq:pipiamp}
\end{equation}
The relevant diagrams are illustrated in Figure \ref{fig:ttwo}; they include a  four-pion contact interaction and also 
$s$, $t$, and $u$-channel exchange of the massive vector boson fields.

Using the pion wavefunction in Eq. (\ref{eq:pionwv}),
we find the 4-pion contact coupling to be of the form
\begin{equation}
  3 g_{4\pi} = \sum_{j=1}^{N+1} \dfrac{v^4}{f_j^6}.
\label{eq:contact}
\end{equation}
Defining the couplings of the pions to the $n$th massive vector boson ($V_n$) by
\begin{equation}
{\cal L}_{V_n \pi \pi }= - g_{V_n\pi\pi} \varepsilon^{abc} \partial_\mu \pi^a\, \pi^b\, V_n^{\mu c}~,
\label{eq:gpipin}
\end{equation}
and using the pion wavefunction,
we compute\footnote{Note that Son and Stephanov used a non-standard normalization
for the nonlinear sigma model kinetic energy terms in Eq. (2.1) of 
\protect\cite{Son:2003et}.} the couplings of the massless pion to be \cite{Son:2003et}
\begin{equation}
g_{V_n\pi\pi}= \sum_{m=1}^{N+1}\frac{v^2}{2\,f^2_m}
\left[ g_{m-1} v^{V_n}_{m-1} + g_m v^{V_n}_m \right]~,
\end{equation}
where $v^{V_n}_m$ is the $m$th element of the (normalized) mass eigenvector corresponding
to the $n$th mass eigenstate $V_n$, which we may write
\begin{equation}
v^{V_n}_m = \langle m | V_n \rangle~.
\end{equation}
We denote the mass of eigenstate $V_n$ as $M_n$.
Recall that $g_0 = g_{N+1} \equiv 0$.  

Combining Eq. (\ref{eq:gpipin}) with Eq. (\ref{eq:pipiamp}), we find
\begin{equation}
  A(s,t,u) = 3 g_{4\pi} s 
            -\sum_n g^2_{V_n\pi\pi} \left[
              \dfrac{u-s}{M_n^2-t}
             +\dfrac{t-s}{M_n^2-u}
             \right],
\label{eq:amplitude}
\end{equation}
with the usual definitions of the kinematic variables
\begin{align}
s & =(p_i + p_j)^2 = 2 p_i \cdot p_j = (p_k+p_l)^2= 2 p_k \cdot p_l~,\\
t & = (p_i-p_k)^2=-2 p_i\cdot p_k = (p_j-p_l)^2=-2 p_j \cdot p_l~,\\
u & = (p_i-p_l)^2 = -2 p_i \cdot p_l = (p_j-p_k)^2=-2 p_j \cdot p_k~,
\end{align}
and $s+t+u \equiv 0$.  Because the contact interaction term gives a contribution
of order ${\cal O}(1/N^2)$, it vanishes in the continuum limit $N \to \infty$
\cite{Son:2003et}.
The only contribution to pion scattering in this limit, therefore, arises from
the exchange of the massive vector bosons. 

We can verify that the model retains the appropriate chiral limit even in the absence
of a four-pion contact term.  The computation is easiest in an $f$-flat deconstruction 
described above, where
\begin{equation}
f^2_m = v^2\, (N+1)~,\ \ \ \ m=1,2,\ldots,N+1~.
\label{eq:fflat}
\end{equation}
We start by noting that 
\begin{equation}
g_m v^{V_n}_m = \langle m| G | V_n \rangle~,
\end{equation}
where $G$ is the $N \times N$ diagonal matrix of gauge couplings with
diagonal elements $(g_1,g_2,\ldots,g_N)$. From Eqs. (\ref{eq:gpipin}) and (\ref{eq:fflat}),
we then find
\begin{equation}
g_{V_n \pi\pi} = \frac{1}{N+1} \sum_{m=1}^N  \langle m | G | V_n \rangle = 
\frac{1}{N+1} \langle \phi | G | V_n\rangle~,
\label{eq:gvnppi}
\end{equation}
where we have introduced the notation
\begin{equation}
| \phi \rangle  = \sum_{m=1}^N | m \rangle~.
\end{equation}
We also make the definition 
\begin{equation}
D(Q^2) = \sum_{n=1}^{N} \frac{g^2_{V_n\pi\pi}}{{Q^2 + {M^2_{n}}}}~,
\label{eq:dqsqd}
\end{equation}
which puts the scattering amplitude into the more compact form
\begin{equation}
A(s,t,u) = (s-u) D(-t)+(s-t) D(-u)~.
\label{eq:astu}
\end{equation}

Combining Eq. (\ref{eq:dqsqd}) with Eq. (\ref{eq:gvnppi}), we can rewrite  $D(Q^2)$ as
\begin{equation}
D(Q^2)  = {1\over (N+1)^2} \sum_{n=1}^N 
{{\langle \phi | G | V_n \rangle \langle V_n | G | \phi \rangle}\over Q^2 + {M}^2_n}~,
\end{equation}
which by completeness of the vectors $|V_n\rangle$ becomes
\begin{equation}
D(Q^2)  = {1\over (N+1)^2}\, \langle \phi | G\, {1\over Q^2 + {\cal M}^2}\, G | \phi \rangle~,
\end{equation} 
where ${\cal M}^2$ is the $KK$ mass matrix: ${\cal M}^2 | V_n \rangle
= M^2_n | V_n\rangle$.
We can then evaluate $D(Q^2=0)$
\begin{equation}
D(0)  = {1\over (N+1)^2}\,  \langle \phi | G\, {1\over {\cal M}^2}\, G | \phi \rangle~, 
 = {1\over (N+1)^2}\, \sum_{i,j=1}^N \left( {4 \over {\cal V}^2}\right)_{i,j}~,
\end{equation}
where ${\cal V}^2$ is the matrix of ``vevs'', ${\cal M}^2 = \frac14 G\, {\cal V}^2\, G$. From Eq.
(B.6) of \cite{SekharChivukula:2004mu}, we find
\begin{equation}
 \left( {4 \over {\cal V}^2}\right)_{i,j} = \frac{4}{v^2 (N+1)^2}\,[N+1 - \max(i,j)]\,\min(i,j)~.
 \end{equation}
 In the continuum limit, $N \to \infty$, we find
 \begin{equation}
 D(0) \to 2 \cdot {4 \over v^2}\, \int^1_0 dx \int^x_0 dy
 (1-x) y = {1\over 3 v^2}~.
 \label{eq:lowenergy}
 \end{equation}
 This yields the correct low-energy theorem
 \begin{equation}
 \lim_{s \to 0} A(s,t,u) = {(2s-u-t) \over 3v^2} = {s\over v^2}~,
 \end{equation}
confirming that this model has the appropriate chiral behavior without having 
a contact interaction.\footnote{As noted by \protect\cite{Harada:1995dc},
a contact term is required if only the $\rho$ meson is included.}

It is useful to note that if we compare the low energy expansion of Eq.(\ref{eq:amplitude}),
\begin{equation}
  A(s,t,u) = 3 g_{4\pi} s - \sum_n \dfrac{g^2_{V_n\pi\pi}}{M_n^2} 
  (t+u-2s)
  +\cdots
  = 3 \left(g_{4\pi} + \sum_n \dfrac{g^2_{V_n\pi\pi}}{M_n^2} \right)s 
  +\cdots,
\end{equation}
with the low energy theorem of $\pi \pi$ scattering  $A(s,t,u) = \dfrac{s}{v^2}$, 
we obtain a sum rule derived by Da Rold and Pomarol \cite{DaRold:2005zs},
\begin{equation}
  \dfrac{1}{v^2} = 3 \left(g_{4\pi} 
    + \sum_n \dfrac{g^2_{V_n\pi\pi}}{M_n^2} \right).
\label{eq:sum_rule1}
\end{equation}
As mentioned above, the four-pion
contact interaction Eq.(\ref{eq:contact}) vanishes in the continuum limit $N\to \infty$ ($f_j\to \infty$).
Eq.(\ref{eq:sum_rule1}) then reads
\begin{equation}
  \dfrac{1}{v^2} = 3 \sum_n \dfrac{g^2_{V_n\pi\pi}}{M_n^2},
\label{eq:sum_rule2}
\end{equation}
which should be compared with the celebrated KSRF 
relation \cite{Kawarabayashi:1966kd,Riazuddin:1966sw} for the $\rho$ meson
in hadron dynamics (but note the different coefficients):
\begin{equation}
  \dfrac{1}{f_\pi^2} = 2 \dfrac{g_{\rho\pi\pi}^2}{M_\rho^2}.
\label{eq:KSRF}
\end{equation}
We will refer to Eq.(\ref{eq:sum_rule2}) as the DP sum 
rule.

\section{Partial Wave Scattering Amplitudes}

We next calculate the partial wave amplitude
\begin{equation}
  T^I_{\ell} = \dfrac{1}{64\pi} \int_{-1}^1 d(\cos\theta)
  P_\ell(\cos\theta) T^{I}(s,t,u), 
  \label{eq:spin0}
\end{equation}
where $\cos\theta$ dependence of $t$ and $u$ are
\begin{equation}
  t = -\frac{s}{2}(1-\cos\theta), \quad
  u = -\frac{s}{2}(1+\cos\theta),
\end{equation}
and $P_\ell$ denotes the Legendre polynomials: $P_0(z) = 1,\, P_1(z) = z, \,
  P_2(z) = \frac{1}{2} (3z^2-1)$.
The isospin decomposition of the pion scattering amplitude $T_\ell^I$
can be performed by
\begin{eqnarray}
  T^{I=0}(s,t,u) &=&  3A(s,t,u) + A(t,s,u) + A(u,t,s), \\
  T^{I=1}(s,t,u) &=&  A(t,s,u) - A(u,t,s), \\
  T^{I=2}(s,t,u) &=&  A(t,s,u) + A(u,t,s).
  \label{eq:azero}
\end{eqnarray}
From the form of Eq. (\ref{eq:amplitude}) we find that $T^{I=0} = - 2 T^{I=2}$.
We obtain
\begin{eqnarray}
  T_{\ell=0}^{I=0}(s)
  &=& \dfrac{3 g_{4\pi} }{16\pi} s
     +\dfrac{1}{8\pi} \sum_n g_{V_n\pi\pi}^2\left[
       2\ln\left(1+\dfrac{s}{M_n^2}\right)
      +\dfrac{M_n^2}{s} \ln\left(1+\dfrac{s}{M_n^2}\right) -1
      \right],
\label{eq:T00}
  \\
  T_{\ell=1}^{I=1}(s)
  &=& \dfrac{g_{4\pi}}{32\pi} s
     +\dfrac{1}{48\pi}\sum_n
      \dfrac{g_{V_n\pi\pi}^2}{1-\dfrac{s}{M_n^2}}
      \Biggl\{
        6\dfrac{M_n^4}{s^2}\left[
          \ln\left(1+\dfrac{s}{M_n^2}\right) - \dfrac{s}{M_n^2} 
         +\dfrac{1}{2}\dfrac{s^2}{M_n^4} 
        \right]
      \Biggr.
  \nonumber\\
  & & \qquad
        +9\dfrac{M_n^2}{s}\left[
          \ln\left(1+\dfrac{s}{M_n^2}\right) - \dfrac{s}{M_n^2} 
        \right]
        -9\ln\left(1+\dfrac{s}{M_n^2}\right)
  \nonumber\\
  & & \qquad
      \Biggl.
        -6\dfrac{s}{M_n^2}\ln\left(1+\dfrac{s}{M_n^2}\right)
        +13\dfrac{s}{M_n^2}
      \Biggr\}~,
\label{eq:T11}
\\
  T_{\ell=0}^{I=2}(s)& =& - \frac{1}{2}   T_{\ell=0}^{I=0}(s)~.
  \label{eq:T20}
\end{eqnarray}
In the low-energy limit, $s \ll M_n^2$, Eqns.(\ref{eq:T00}), (\ref{eq:T11}) and (\ref{eq:T20}) can be
expanded as
\begin{eqnarray}
  T_0^0(s) &=& -2 T_0^2(s) = \dfrac{3g_{4\pi}}{16\pi} s +
  \dfrac{3}{16\pi} \sum_n \dfrac{g_{V_n\pi\pi}^2}{M_n^2} s
  \left(
    1 - \frac{4}{9}\dfrac{s}{M_n^2} + \cdots
  \right),
\label{eq:T00low}
  \\
  T_1^1(s) &=& \dfrac{g_{4\pi}}{32\pi} s +
  \dfrac{1}{32\pi}\sum_n \dfrac{g_{V_n\pi\pi}^2}{M_n^2} s
  \left(
    1 + \dfrac{s}{M_n^2} + \cdots
  \right).
\label{eq:T11low}
\end{eqnarray}
Conversely, the asymptotic form of $T^0_0$ at high energies is
\begin{equation}
 \lim_{s\to \infty} T^0_0(s) = \frac{1}{4\pi} \sum_{n} g_{V_n\pi\pi}^2 \ln\frac{s}{M^2_{1}} + ...~,
 \end{equation}
where we anticipate that the couplings $g^2_{V_n \pi \pi}$ will fall sufficiently fast
to allow the expression above to converge.
Evaluating the limiting forms of the scattering amplitudes therefore involves the sums
\begin{equation}
  \sum_{n} g_{V_n\pi\pi}^2, \qquad
  \sum_{n} \dfrac{g_{V_n\pi\pi}^2}{M_n^2}, \qquad
  \sum_{n} \dfrac{g_{V_n\pi\pi}^2}{M_n^4}, 
\label{eq:sums}
\end{equation}
the middle of which is given by the DP sum rule.  We will shortly calculate these sums in a range of global linear moose models.   

Before beginning that discussion, however, note that 
the pion scattering amplitude can also be calculated from the
electroweak chiral Lagrangian \cite{Boos:1997gw},
\begin{equation}
  A(s,t,u) = \dfrac{s}{v^2} 
     + \alpha_4 \dfrac{4(t^2 + u^2)}{v^4}
     + \alpha_5 \dfrac{8s^2}{v^4}.
\end{equation}
The corresponding partial wave amplitudes are
\begin{eqnarray}
  T_0^0(s) &=& \dfrac{1}{16\pi}\dfrac{s}{v^2}
   \left[ 1 + \dfrac{4}{3}(7\alpha_4 + 11\alpha_5) \dfrac{s}{v^2}
   \right], \label{eq:lont1} \\
  T_1^1(s) &=& \dfrac{1}{96\pi}\dfrac{s}{v^2}
   \left[ 1 + (4\alpha_4 - 8\alpha_5) \dfrac{s}{v^2}
   \right]~,\\
   T_0^2(s) &=& \dfrac{-1}{32\pi}\dfrac{s}{v^2} 
   \left[ 1 - \dfrac{16}{3} (2\alpha_4 + \alpha_5)\dfrac{s}{v^2}
   \right]. \label{eq:lont3}
\label{eq:chpt_amp}
\end{eqnarray}
Comparing these expressions to Eqs. (\ref{eq:T00low}) and (\ref{eq:T11low}),
we immediately find the additional relation
\begin{equation}
\alpha_4 = -\alpha_5 = {v^4\over 4} \sum_n {g^2_{V_n \pi \pi} \over M^4_n}~.
\end{equation}
which yields the relation $T^0_0(s) = - 2 T^2_0(s)$ again when applied to Eqns. (\ref{eq:lont1})-(\ref{eq:lont3}).
Using the results of the next section therefore allows us to reproduce the
Longhitano parameters calculated for various linear Moose models in \cite{Chivukula:2005ji}.


\section{Applications: $\pi\pi$ elastic scattering in particular Moose models}

We now calculate the sums in Eq. (\ref{eq:sums}) and the resulting partial wave elastic $\pi\pi$ scattering amplitudes in a range of global linear moose models, starting with a flat deconstruction and then introducing a more general language that facilitates calculation in the warped case.

\subsection{The flat case}
\label{flatcase}

We first consider a flat deconstruction lattice
\begin{equation}
  g_1 = g_2 = \cdots = g_N = \sqrt{N} \tilde{g}, \qquad
  f_1 = f_2 = \cdots = f_{N+1} = \sqrt{N+1} v,
\end{equation}
and the continuum limit ($N\to \infty$) of the 5D action,
\begin{equation}
  S = \int_0^{\pi R} dy \int d^4 x \dfrac{1}{g_5^2} \left\{
    -\dfrac{1}{4} F^a_{MN} F^{aMN} \right\}.
\label{eq:continuum}
\end{equation}
The gauge field $A_\mu$ satisfies Dirichlet boundary conditions 
$A_\mu(x,y=0) = A_\mu(x,y=\pi R) = 0$, while $A_y$ satisfies Neumann boundary conditions $\partial_y (A_y(x,y=0)) = \partial_y(A_y(x,y=\pi R))=0$. 
Latticizing the continuum action Eq.(\ref{eq:continuum}) we see
\begin{equation}
  \dfrac{\pi R}{N} \dfrac{1}{g_5^2} = \dfrac{1}{N \tilde{g}^2}, \qquad
  \dfrac{\pi R}{N} = \dfrac{2}{\sqrt{N(N+1)} \tilde{g} v},
\end{equation}
with $\pi R/N$ being the lattice spacing.  We thus obtain
\begin{equation}
  \dfrac{\pi R}{g_5^2} = \dfrac{1}{\tilde{g}^2}, \qquad
  \pi R = \dfrac{2}{\tilde{g} v},
\label{eq:dic}
\end{equation}
in the continuum limit.
Eq.(\ref{eq:dic}) plays the role of a dictionary between the
deconstruction and continuum languages $(\tilde{g},v) \leftrightarrow (g_5,R)$.
For example, the $n$th vector boson mass can be expressed as
\begin{equation}
  M_n^2 = \dfrac{n^2}{R^2} = \dfrac{n^2 \pi^2 \tilde{g}^2 v^2}{4}.
\label{eq:Vnmass}
\end{equation}

The mode-functions of the pion and the $n$th vector boson are given by
\begin{equation}
  \chi_{(\pi)}(y) = \dfrac{g_5}{\sqrt{\pi R}},
\hspace{2cm}
  \chi_{(Vn)}(y) = \sqrt{\dfrac{2}{\pi R}} 
     g_5 \sin\left(\dfrac{ny}{R}\right),
\end{equation}
which satisfy the normalization conditions,
\begin{equation}
  1=\dfrac{1}{g_5^2} \int_0^{\pi R} dy |\chi_{(\pi)}(y)|^2, \qquad
  \delta_{nm}
   =\dfrac{1}{g_5^2} \int_0^{\pi R} dy \chi_{(Vn)}(y) \chi_{(Vm)}(y).
\end{equation}
Now it is straightforward to calculate the $V_n \pi\pi$ vertex,
\begin{equation}
  g_{V_n \pi\pi} = \dfrac{1}{g_5^2} \int_0^{\pi R} dy \chi_{(Vn)}(y)
  |\chi_{(\pi)}(y)|^2, 
\end{equation}
and we obtain
\begin{equation}
  g_{V_n\pi\pi}^2 = \left\{
    \begin{array}{ll}
      \dfrac{g_5^2}{\pi R} \dfrac{8}{n^2\pi^2} & \mbox{ for $n$ : odd}
      \\
      0 & \mbox{ for $n$ : even}
    \end{array}
  \right. \hspace{1cm}   = \left\{
    \begin{array}{ll}
      \dfrac{8 \tilde{g}^2}{n^2\pi^2} & \mbox{ for $n$ : odd}
      \\
      0 & \mbox{ for $n$ : even}
    \end{array}
  \right.
  \label{eq:Vpipi}
\end{equation}
in the continuum and deconstruction languages, respectively.
Combining Eq.(\ref{eq:Vnmass}) and Eq.(\ref{eq:Vpipi}) we obtain
\begin{equation}
  \dfrac{g_{V_n\pi\pi}^2}{M_n^2} = \left\{
    \begin{array}{ll}
      \dfrac{1}{v^2}\dfrac{32}{n^4\pi^4} & \mbox{ for $n$ : odd}
      \\
      0 & \mbox{ for $n$ : even}
    \end{array}
  \right. \hspace{1cm}
   \dfrac{g_{V_n\pi\pi}^2}{M_n^4} = \left\{
    \begin{array}{ll}
      \dfrac{1}{M_1^2 v^2}\dfrac{32}{n^6\pi^4} & \mbox{ for $n$ : odd}
      \\
      0 & \mbox{ for $n$ : even}.
    \end{array}
  \right.
  \end{equation}
  
 We may now evaluate the sums involved in the limiting forms of the scattering amplitudes.   First, the sum over squared couplings:
 \begin{equation} 
  \sum_n {g_{V_n\pi\pi}^2}
  =  \tilde{g}^2 \sum_{j=0}^{\infty} \dfrac{8}{(2j+1)^2 \pi^2} = \tilde{g}^2.
\end{equation}
Next, we verify the DP sum rule:
\begin{equation}
  3\sum_n \dfrac{g_{V_n\pi\pi}^2}{M_n^2}
  = \dfrac{1}{v^2} \sum_{j=0}^{\infty} \dfrac{96}{(2j+1)^4 \pi^4}
  = \dfrac{1}{v^2}\ ,
\end{equation}
noting that  the first vector resonance almost
saturates the DP sum rule by about $98.6$\% ($96/\pi^4 = 0.986...$).
Third, the sum with inverse fourth powers of KK masses:
\begin{equation}
  3\sum_n \dfrac{g_{V_n\pi\pi}^2}{M_n^4}
    = \dfrac{1}{v^2M_1^2} \sum_{j=0}^{\infty} \dfrac{96}{(2j+1)^6 \pi^4}
  = \dfrac{\pi^2}{10}\dfrac{1}{v^2 M_1^2}.
\end{equation}
The low energy partial wave amplitudes
Eqs.(\ref{eq:T00low},\ref{eq:T11low}) are then
\begin{eqnarray}
  T_0^0(s) &=& - 2 T_0^2(s) = \dfrac{1}{16\pi} \dfrac{s}{v^2}\left[
    1 - \dfrac{2\pi^2}{45} \dfrac{s}{M_1^2} + \cdots \right],
\label{eq:flat_amp0}
  \\
  T_1^1(s) &=& \dfrac{1}{96\pi} \dfrac{s}{v^2}\left[
    1 + \dfrac{\pi^2}{10} \dfrac{s}{M_1^2} + \cdots \right],
\label{eq:flat_amp1}
\end{eqnarray}
while the high-energy asymptotic partial wave amplitude is
\begin{equation}
 \lim_{s\to \infty} T^0_0(s) = \frac{\tilde{g}^2}{4\pi} \ln\frac{s}{M^2_{1}}+\ldots~.
 \label{eq:asymp00}
\end{equation}

\subsection{The general case}

If we, instead, deconstruct an arbitrary Higgsless model
in conformal coordinates\footnote{That is, for a deconstructed model we block spin such that
Eq. (\protect{\ref{eq:conformalcond}}) holds.} 
 the continuum can be described by a 5D action (here we absorb factors of $g_5$ in the function
$\kappa(z)$)
\begin{equation}
  S = \int_{R_1}^{R_2} dz \,\kappa(z) \int d^4 x \left[
    -\frac{1}{4} \eta^{\mu\alpha}\eta^{\nu\beta} 
     F^a_{\mu\nu} F^a_{\alpha\beta}
    +\frac{1}{2} \eta^{\mu\nu}
     F^a_{\mu z} F^a_{\nu z}
  \right],
\label{eq:general_action}
\end{equation}
where $F_{MN}$ is the field strength of the 5D gauge field $A_M(x,z)$.
The gauge field $A_\mu$ satisfies Dirichlet boundary conditions at both boundaries:
$A_\mu(x,z=R_1) = A_\mu(x,z=R_2) = 0$.
The arbitrary function $\kappa(z)$ encodes (in conformal coordinates) 
the warped extra dimension or the
position dependent gauge couplings.  In this section, we show how to calculate the
sums in Eq. (\ref{eq:sums}) 
and the related partial wave scattering amplitudes solely from $\kappa(z)$, without knowing the  detailed form of the wavefunctions of the massive KK modes $V_n$.

We start with some general information about the wavefunctions.
The mode-functions of the massive KK-modes of $A_\mu$ obey
\begin{equation}
  0 = \dfrac{d}{dz} \left[ 
    \kappa(z) \dfrac{d}{dz} \chi_{(Vn)}(z) 
  \right] + \kappa(z) M_n^2 \chi_{(Vn)}(z).
\label{eq:mode_eq}
\end{equation}
These mode-functions satisfy the orthonormality and completeness relations
\cite{Hirn:2005nr,Sakai:2005yt}
\begin{equation}
  \delta_{n,m} = \int_{R_1}^{R_2} dz \kappa(z) 
    \chi_{(Vn)}(z) \chi_{(Vm)}(z) \qquad   \delta(z-z') = 
  \kappa(z) \sum_n \chi_{(Vn)}(z) \chi_{(Vn)}(z')  .
\label{eq:completeness}
\end{equation}
There also exists an uneaten Nambu-Goldstone degree of freedom (the pion) in
$A_z$.  Its mode-function and the associated normalization condition are given by
\begin{equation}
  \chi_{(\pi)}(z) \propto \dfrac{1}{\kappa(z)}, \qquad 1 = \int_{R_1}^{R_2} dz \kappa(z) |\chi_{(\pi)}(z)|^2 .
\label{eq:pi_modefunc}
\end{equation}

From the mode-functions of the pion and KK modes, we can calculate the $V_n\pi\pi$ coupling  
\begin{equation}
  g_{V_n\pi\pi} 
  = \int_{R_1}^{R_2} dz \kappa(z) |\chi_{(\pi)}(z)|^2 \chi_{(Vn)}(z)
  \nonumber\\
  = \dfrac{\int_{R_1}^{R_2} dz \kappa^{-1}(z) \chi_{(Vn)}(z)}
            {\int_{R_1}^{R_2} dz \kappa^{-1}(z)}.
\end{equation}
Using the convenient definition 
\begin{equation}
  \Delta^{(i)}(z,z') \equiv 
  \sum_n \chi_{(Vn)}(z) \dfrac{1}{M_n^{2i}} \chi_{(Vn)}(z')
\end{equation}
allows us to write the sums of Eq. (\ref{eq:sums}) in the compact form
\begin{equation}
  \sum_n \dfrac{g_{V_n\pi\pi}^2}{M_n^{2i}}
  =  \dfrac{
        \int_{R_1}^{R_2} dz \int_{R_1}^{R_2} dz' 
        \kappa^{-1}(z) \kappa^{-1}(z') \Delta^{(i)}(z,z')
     }{
        \left[\int_{R_1}^{R_2} dz \kappa^{-1}(z)\right]^2
     }.
\label{eq:sum_general}
\end{equation}
Because the completeness relation (\ref{eq:completeness}) tells us 
\begin{equation}
  \Delta^{(0)}(z,z') = \dfrac{1}{\kappa(z)} \delta(z-z'),
\end{equation}
we see that we can calculate the $i=0$ sum, $\sum_n g_{V_n\pi\pi}^2$ from $\kappa(z)$, without
knowing the detailed form of the mode-function $\chi_{Vn}(z)$:
\begin{equation}
  \sum_n g_{V_n\pi\pi}^2
  =  \dfrac{
        \int_{R_1}^{R_2} dz \kappa^{-3}(z)
     }{
        \left[\int_{R_1}^{R_2} dz \kappa^{-1}(z)\right]^2
     }.
\label{eq:sum_rule0}
\end{equation}

A little more work yields similar results for the other sums in Eq. (\ref{eq:sums}).
The mode-equation Eq.(\ref{eq:mode_eq}) leads to
\begin{equation}
  \dfrac{d}{dz} \left[ 
    \kappa(z) \dfrac{d}{dz} \Delta^{(i)}(z,z')
  \right] = -\kappa(z) \Delta^{(i-1)}(z,z').
\label{eq:diff_eq_delta}
\end{equation}
Combining Eq.(\ref{eq:diff_eq_delta}) and the Dirichlet boundary
conditions 
\begin{equation}
  \Delta^{(i)}(z=R_1,z') = \Delta^{(i)}(z=R_2,z') = 0
\end{equation}
we obtain
\begin{eqnarray}
  \Delta^{(1)}(z,z') 
  &=& \dfrac{K(R_1)}{K(R_2)-K(R_1)}\left[
        K(z)\theta(z-z') + K(z') \theta(z'-z)
      \right]
  \nonumber\\
  & & 
     +\dfrac{K(R_2)}{K(R_2)-K(R_1)}\left[
        K(z)\theta(z'-z) + K(z') \theta(z-z')
      \right] 
  \nonumber\\
  & &
     -\dfrac{1}{K(R_2)-K(R_1)}\left[
        K(R_1) K(R_2) + K(z) K(z')
      \right], 
\end{eqnarray}
with $K(z)$ defined by
\begin{equation}
  \dfrac{d}{dz} K(z) = \kappa^{-1}(z).
\label{eq:def_K(z)}
\end{equation}
The higher $\Delta^{(i)}$ can then be calculated recursively 
\begin{equation}
  \Delta^{(i)}(z,z')
  = \int_{R_1}^{R_2} dz'' \Delta^{(1)}(z,z'') \kappa(z'') 
    \Delta^{(i-1)}(z'',z').
\end{equation}
Combining these expressions with Eq.(\ref{eq:sum_general})
we find
\begin{equation}
  \sum_n \dfrac{g_{V_n\pi\pi}^2}{M_n^2}
  = \dfrac{
        \int_{R_1}^{R_2} dz_1 
        \int_{R_1}^{R_2} dz_2
        \kappa^{-1}(z_1) \Delta^{(1)}(z_1,z_2)
        \kappa^{-1}(z_2)
     }{
        \left[\int_{R_1}^{R_2} dz \kappa^{-1}(z)\right]^2
     },
\label{eq:sum_rule1a}
\end{equation}
and
\begin{equation}
  \sum_n \dfrac{g_{V_n\pi\pi}^2}{M_n^4}
  = \dfrac{
        \int_{R_1}^{R_2} dz_1 
        \int_{R_1}^{R_2} dz_2
        \int_{R_1}^{R_2} dz_3
        \kappa^{-1}(z_1) \Delta^{(1)}(z_1,z_2)
        \kappa(z_2) \Delta^{(1)}(z_2,z_3) \kappa^{-1}(z_3) 
     }{
        \left[\int_{R_1}^{R_2} dz \kappa^{-1}(z)\right]^2
     }.
\label{eq:sum_rule2a}
\end{equation}
As promised, we can calculate all of the sums in Eq. (\ref{eq:sums}) directly from $\kappa(z)$.  We now apply this to several global linear moose models.

\subsubsection{Application to $SU(2)$ warped case}

Let us apply the general method to an $SU(2)$ linear moose model in warped space. In conformal
coordinates, with a metric given by Eq. (\ref{eq:conformalmetric}), the continuum limit of this model 
is described by the 5D action\footnote{For notational convenience, the coordinate $z$ here is
dimensionful and  corresponds to $Z$ in Eq. (\protect{\ref{eq:conformalmetric}}).}
\begin{equation}
  S = \int_{R}^{R'} dz \dfrac{1}{z \tilde{g}_5^2} 
      \int d^4 x \left[
        -\frac{1}{4}\eta^{\mu\alpha}\eta^{\nu\beta} 
          F^a_{\mu\nu} F^a_{\alpha\beta}
        +\frac{1}{2} \eta^{\mu\nu} F^a_{\mu z} F^a_{\nu z}
      \right].
      \label{eq:su2warpedlag}
\end{equation}
Here $\tilde{g}^2_5=g^2_5/R$ is dimensionless, where $g_5$ is the usual 5-dimensional coupling,
and $A_\mu$ satisfies Dirichlet boundary conditions at both
boundaries $z=R, R'$ (corresponding to the global linear moose).
This Lagrangian is of
the form of Eq.(\ref{eq:general_action}) with 
\begin{equation}
\kappa(z) = \frac{1}{\tilde{g}_5^2 z}\qquad\qquad R_1 = R, \,\,R_2 = R'
\end{equation}
and we may solve Eq. (\ref{eq:def_K(z)}) to find 
\begin{equation}
K(z) = \frac{1}{2}\tilde{g}^2_5 z^2.
\end{equation}

From this point on, we assume a large hierarchy between $R$
and $R'$,
\begin{equation}
  R = R'\exp(-b/2) \ll R' .
\end{equation}
The integrals of Eq.(\ref{eq:sum_rule0}), Eq.(\ref{eq:sum_rule1a}) and Eq.(\ref{eq:sum_rule2a}) 
then yield the results
\begin{eqnarray}
  \sum_n g_{V_n\pi\pi}^2
  &=& \tilde{g}_5^2,
\label{eq:sum0a} 
  \\
  \sum_n \dfrac{g_{V_n\pi\pi}^2}{M_n^2}
  &=& \dfrac{1}{24} (R')^2 \tilde{g}_5^2, 
\label{eq:sum1a} 
  \\
  \sum_n \dfrac{g_{V_n\pi\pi}^2}{M_n^4}
  &=& \dfrac{1}{384} (R')^4 \tilde{g}_5^2. 
\label{eq:sum2a} 
\end{eqnarray}
Recalling that Eq. (\ref{eq:sum1a}) must satisfy the DP sum rule and using
\begin{equation}
M_1 = \dfrac{x_1}{R'}, \qquad
  x_1 \simeq 3.8317,
\end{equation}
Eqs.(\ref{eq:sum0a}), (\ref{eq:sum1a}) and (\ref{eq:sum2a}) can be
rewritten as
\begin{eqnarray}
  \sum_n g_{V_n\pi\pi}^2
  &=& \dfrac{8 M_1^2}{x_1^2 v^2},
\label{eq:sum0b} 
  \\
  \sum_n \dfrac{g_{V_n\pi\pi}^2}{M_n^2}
  &=& \dfrac{1}{3v^2},
\label{eq:sum1b} 
  \\
  \sum_n \dfrac{g_{V_n\pi\pi}^2}{M_n^4}
  &=& \dfrac{1}{48 v^2} \dfrac{x_1^2}{M_1^2}.
\label{eq:sum2b} 
\end{eqnarray}
The low energy partial wave amplitudes
Eqs.(\ref{eq:T00low}) and (\ref{eq:T11low}) are then
\begin{eqnarray}
  T_0^0(s) &=& - 2 T_0^2(s) = \dfrac{1}{16\pi} \dfrac{s}{v^2}\left[
    1 - \dfrac{x_1^2}{36} \dfrac{s}{M_1^2} + \cdots \right],
\label{eq:warped_amp0}
  \\
  T_1^1(s) &=& \dfrac{1}{96\pi} \dfrac{s}{v^2}\left[
    1 + \dfrac{x_1^2}{16} \dfrac{s}{M_1^2} + \cdots \right].
\label{eq:warped_amp1}
\end{eqnarray}

Numerical analysis shows that the contributions from the first KK mode
 to the sums above are substantial: 
$54.4$\% for Eq.(\ref{eq:sum0a}), $89.1$\% for
Eq.(\ref{eq:sum1a}), and $97.1$\% for Eq.(\ref{eq:sum2a}).
The contributions from the first ten KK-modes are:
$92.5$\% for Eq.(\ref{eq:sum0a}), $99.9$\% for
Eq.(\ref{eq:sum1a}), and almost 100\% for Eq.(\ref{eq:sum2a}).
The convergence of Eq.(\ref{eq:sum0a}) is quite slow in this $SU(2)$  model, as
compared with other models examined in this paper, because the model lacks 
 a parity symmetry.  In a parity-symmetric model, the second-largest
contribution to the sum in Eq. (\ref{eq:sum0a}) arises from the n=3 mode which has
a far smaller coupling to pions than the $n=1$ mode; in the $SU(2)$ model, the second-largest contribution is from the $n=2$ mode, whose coupling to pions is not as heavily suppressed.
Sums (\ref{eq:sum1a}) and (\ref{eq:sum2a}) still converge rapidly in this model because the hierarchy of KK masses is substantial.

\TABLE[ht]
{ \begin{tabular}{|c|cccccc|}
    \hline
    $n$
      & 1 & 2 & 3 & 4 & 5 & 6 \\
    \hline
    $\phantom{\dfrac{\strut}{\strut}}M_n R'$ 
      & $3.831706$ & $7.015587$ & $10.17347$ & $13.32369$ &
        $16.47063$ & $19.61586$ \\
    \hline
    $\dfrac{g_{V_n\pi\pi}^2}{\tilde{g}_5^2}$
      & $0.544886$ & $0.162541$ & $0.077295$ & $0.045065$ &
        $0.029490$ & $0.020791$ \\
    \hline
  \end{tabular}
  \caption{Numerical results for $M_n$ and $g_{V_n\pi\pi}^2$ in the $SU(2)$ warped-space model.}
  \label{tab:numericala}}

\subsubsection{Application to $SU(2)_L \times SU(2)_R$ flat case}

We now apply the general method to an $SU(2)\times SU(2)$ model in flat space.
The 5D action of this model is given by
\begin{eqnarray}
  S &=&
    \int_0^{\pi R} dz \int d^4 x \left\{
    \dfrac{1}{g_{5L}^2}\left[
        -\frac{1}{4} \eta^{\mu\alpha}\eta^{\nu\beta} 
         L^a_{\mu\nu} L^a_{\alpha\beta}
        +\frac{1}{2} \eta^{\mu\nu}
         L^a_{\mu z} L^a_{\nu z}
    \right]
    \right.
  \nonumber\\
    & & \qquad\qquad\qquad
    \left. 
   +\dfrac{1}{g_{5R}^2}\left[
        -\frac{1}{4} \eta^{\mu\alpha}\eta^{\nu\beta} 
         R^a_{\mu\nu} R^a_{\alpha\beta}
        +\frac{1}{2} \eta^{\mu\nu}
         R^a_{\mu z} R^a_{\nu z}
    \right]
   \right\},
\label{eq:flat_action}
\end{eqnarray}
where the $SU(2)\times SU(2)$ gauge fields $L_\mu$ and $R_\mu$ satisfy
the boundary conditions, 
\begin{equation}
  L_\mu(x,z=0) = R_\mu(x,z=0) = 0, \qquad
  L_\mu(x,z=\pi R) - R_\mu(x,z=\pi R) = 0 .
\end{equation}
If we make the identifications 
\begin{eqnarray}
  A_\mu(x,z) = L_\mu(x,z),
  & &
  A_z(x,z) = L_z(x,z), 
  \nonumber\\
  A_\mu(x,z+\pi R) = R_\mu(x,z), 
  & &
  A_z(x,z+\pi R) = -R_z(x,z),
\label{eq:flat_identification}
\end{eqnarray}
we find the Lagrangian Eq.(\ref{eq:flat_action}) can be rewritten in 
the form of Eq.(\ref{eq:general_action}) with
\begin{equation}
  \kappa(z) = \dfrac{1}{g_{5L}^2} \theta(\pi R - z)
             +\dfrac{1}{g_{5R}^2} \theta(z - \pi R), \qquad  R_1 = 0, \quad R_2 = 2\pi R .
\end{equation}

With $\kappa(z)$ in hand, we can perform the integrals of Eq.(\ref{eq:sum_rule0}) to obtain
\begin{equation}
  \sum_n g_{V_n\pi\pi}^2
  = \dfrac{1}{\pi R} 
    \dfrac{g_{5L}^4 - g_{5L}^2 g_{5R}^2 + g_{5R}^4}
          {g_{5L}^2 + g_{5R}^2}~,
\label{eq:flat_sum0}
\end{equation}
and solve Eq.(\ref{eq:def_K(z)}) to find
\begin{equation}
  K(z) = g_{5L}^2 (z-\pi R) \theta(\pi R - z)
        +g_{5R}^2 (z-\pi R) \theta(z - \pi R).
\end{equation}
The calculations of Eq.(\ref{eq:sum_rule1}) and Eq.(\ref{eq:sum_rule2})
are a bit  more involved, but yield 
\begin{equation}
  \sum_n \dfrac{g_{V_n\pi\pi}^2}{M_n^2}
  = \dfrac{\pi R}{12} \left(g_{5L}^2 + g_{5R}^2 \right),
\label{eq:flat_sum1}
\end{equation}
\begin{equation}
  \sum_n \dfrac{g_{V_n\pi\pi}^2}{M_n^4}
  = \dfrac{(\pi R)^3}{120}
    \dfrac{g_{5L}^4 + 14 g_{5L}^2 g_{5R}^2 + g_{5R}^4}
          {g_{5L}^2 + g_{5R}^2}.
\label{eq:flat_sum2}
\end{equation}
Note that, if we apply Eq. (\ref{eq:dic}), then Eq.(\ref{eq:flat_sum1}) is consistent with the DP sum
rule, 
\begin{equation}
  \sum_n \dfrac{g_{V_n\pi\pi}^2}{M_n^2} = \dfrac{1}{3v^2}, \qquad
  \dfrac{1}{v^2} 
  = \dfrac{1}{v_L^2} + \dfrac{1}{v_R^2}
  = \dfrac{\pi R}{4} \left(g_{5L}^2 + g_{5R}^2 \right).
\label{eq:ksrf_flat}
\end{equation}
Combining Eq.(\ref{eq:flat_sum0}) and Eq.(\ref{eq:ksrf_flat}) we
obtain
\begin{equation}
  \sum_n g_{V_n\pi\pi}^2 = \dfrac{4}{(\pi R)^2 v^2} 
  \dfrac{g_{5L}^4 - g_{5L}^2 g_{5R}^2 + g_{5R}^4}
        {(g_{5L}^2 + g_{5R}^2)^2}.
\end{equation}

To make contact with our previous flat-space calculation for an $SU(2)$ model,
we now calculate the spectrum and the couplings of spin-1
KK-modes in the parity symmetric limit $g_{5L} = g_{5R} = g_5$:
\begin{equation}
  M_n^2 = \dfrac{n^2}{(2R)^2}, 
  \label{eq:mfparsym22}
\end{equation}
and
\begin{equation}
  g_{V_n\pi\pi}^2 = \left\{
    \begin{array}{ll}
      \dfrac{g_5^2}{\pi R} \dfrac{4}{n^2\pi^2} & \mbox{ for $n$ : odd},
      \\
      0 & \mbox{ for $n$ : even}.
    \end{array}
  \right.
  \label{eq:gfparsym22}
\end{equation}
Note that the $V_n\pi\pi$ couplings of the axial-vector bosons are now
forbidden by the parity invariance, matching the results of section \ref{flatcase}.
We also see that the results of Eqs. (\ref{eq:Vnmass}, \ref{eq:Vpipi}) and Eqs. (\ref{eq:mfparsym22}, \ref{eq:gfparsym22}) are equivalent if we
recall that the parity-symmetric limit of the $SU(2)_L \times SU(2)_R$
corresponds to an interval ($R$) twice as large as that of the $SU(2)$ model. 

The sums Eq.(\ref{eq:flat_sum0}), (\ref{eq:flat_sum1}) 
and (\ref{eq:flat_sum2}) are almost saturated by the first few vector 
resonances; the first KK mode saturates Eq.(\ref{eq:flat_sum0}) about
$81.06$\%, Eq.(\ref{eq:flat_sum1}) about $98.55$\%, and
Eq.(\ref{eq:flat_sum2}) about $99.86$\%.  (
$8/\pi^2  \simeq 0.8106$,
$96/\pi^4 \simeq 0.9855$, and
$960/\pi^6 \simeq 0.9986$.)

\subsubsection{Application to $SU(2)_L \times SU(2)_R$ warped case}

Finally, we consider the $SU(2)\times SU(2)$ model in warped space.  In analogy
to Eq. (\ref{eq:su2warpedlag}), the 5D action is written as 
\begin{eqnarray}
  S &=&
    \int_R^{R'} dz \int d^4 x \left\{
    \dfrac{1}{z \tilde{g}_{5L}^2}\left[
        -\frac{1}{4} \eta^{\mu\alpha}\eta^{\nu\beta} 
         L^a_{\mu\nu} L^a_{\alpha\beta}
        +\frac{1}{2} \eta^{\mu\nu}
         L^a_{\mu z} L^a_{\nu z}
    \right]
    \right.
  \nonumber\\
    & & \qquad\qquad\qquad
    \left. 
   +\dfrac{1}{z \tilde{g}_{5R}^2}\left[
        -\frac{1}{4} \eta^{\mu\alpha}\eta^{\nu\beta} 
         R^a_{\mu\nu} R^a_{\alpha\beta}
        +\frac{1}{2} \eta^{\mu\nu}
         R^a_{\mu z} R^a_{\nu z}
    \right]
   \right\},
\end{eqnarray}
where the gauge fields are assumed to satisfy the boundary conditions
\begin{equation}
  L_\mu(x,z=R) = R_\mu(x,z=R) = 0, \qquad
  L_\mu(x,z=R') - R_\mu(x,z=R') = 0 .
\end{equation}
Field identifications similar to Eq.(\ref{eq:flat_identification})
put the model in the form of Eq.(\ref{eq:general_action}). 
We find
\begin{equation}
  \kappa(z) = \dfrac{1}{z \tilde{g}_{5L}^2} \theta(R'-z)
             +\dfrac{1}{(2R'-z) \tilde{g}_{5R}^2} \theta(z-R'),
\qquad
  R_1 = R, \quad
  R_2 = 2R'-R .
\end{equation}
Knowing this, we may solve Eq.(\ref{eq:def_K(z)}) to find
\begin{equation}
  K(z) = \frac{1}{2} \tilde{g}_{5L}^2 
         \left( z^2 - (R')^2 \right) \theta(R'-z)
        -\frac{1}{2} \tilde{g}_{5R}^2 
         \left( (2R'-z)^2 - (R')^2 \right) \theta(z-R').
\end{equation}

From this point on, we assume a large hierarchy between $R$
and $R'$,
\begin{equation}
  R = R'\exp(-b/2) \ll R' .
\end{equation}
The integrals of Eq.(\ref{eq:sum_rule0}) are straightforward, and those of Eq.(\ref{eq:sum_rule1a}) and Eq.(\ref{eq:sum_rule2a}) a bit less so.  The results are
\begin{eqnarray}
  \sum_n g_{V_n\pi\pi}^2
  &=& \dfrac{\tilde{g}_{5L}^4 - \tilde{g}_{5L}^2 \tilde{g}_{5R}^2 
             + \tilde{g}_{5R}^4}
          {\tilde{g}_{5L}^2 + \tilde{g}_{5R}^2} ~,
\label{eq:warped_sum0}
\\
  \sum_n \dfrac{g_{V_n\pi\pi}^2}{M_n^2}
  &=& \dfrac{(R')^2}{24} \left(\tilde{g}_{5L}^2 + \tilde{g}_{5R}^2 \right),
\label{eq:warped_sum1}
\\
  \sum_n \dfrac{g_{V_n\pi\pi}^2}{M_n^4}
 & = &\dfrac{(R')^4}{384}
    \dfrac{\tilde{g}_{5L}^4 + 9 \tilde{g}_{5L}^2 \tilde{g}_{5R}^2 
             + \tilde{g}_{5R}^4}
          {\tilde{g}_{5L}^2 + \tilde{g}_{5R}^2}.
\label{eq:warped_sum2}
\end{eqnarray}
By using a relation
\begin{equation}
  \dfrac{1}{v^2} = \dfrac{1}{v_L^2} + \dfrac{1}{v_R^2}
    = \dfrac{(R')^2}{8} (\tilde{g}_{5L}^2 + \tilde{g}_{5R}^2 ),
\end{equation}
Eq.(\ref{eq:warped_sum0}) can be rewritten as
\begin{equation}
  \sum_n g_{V_n\pi\pi}^2
  = \dfrac{8}{(R')^2 v^2} 
    \dfrac{\tilde{g}_{5L}^4 - \tilde{g}_{5L}^2 \tilde{g}_{5R}^2 
         + \tilde{g}_{5R}^4}
          {(\tilde{g}_{5L}^2 + \tilde{g}_{5R}^2)^2}.
\end{equation}

The spectrum and the couplings of the first several KK-modes of the
parity symmetric limit $\tilde{g}_{5L} = \tilde{g}_{5R} = \tilde{g}_{5}$ of the warped
$SU(2)\times SU(2)$ model are shown in
Table~\ref{tab:numerical}. Note that the masses of the $n=2j$ modes of this
model match the masses of the $n=j$ modes of the warped-space $SU(2)$ model.
We find the first spin-1 KK-mode saturates Eq.(\ref{eq:warped_sum0})
about $95.68$\%, Eq.(\ref{eq:warped_sum1}) about $99.27$\%, and
Eq.(\ref{eq:warped_sum2}) about $99.87$\%.   

\TABLE[ht]
{  \begin{tabular}{|c|cccccc|}
    \hline
    $n$
      & 1 & 2 & 3 & 4 & 5 & 6 \\
    \hline
    $\phantom{\dfrac{\strut}{\strut}}M_n R'$ 
      & $2.404826$ & 
        $3.831706$ 
      & $5.520078$ &
        $7.015587$ 
      & $8.653728$ &
        $10.17347$ \\
    \hline
    $\dfrac{g_{V_n\pi\pi}^2}{\tilde{g}_5^2}$
      & $0.478394$ &
        $0$
      & $0.017232$ &
        $0$
      & $0.002853$ &
        $0$ \\
    \hline
  \end{tabular}
  \caption{Numerical results of $M_n$ and $g_{V_n\pi\pi}^2$ in
    warped $SU(2)\times SU(2)$ model when 
    $\tilde{g}_5 = \tilde{g}_{5L} = \tilde{g}_{5R}$ is assumed.
    }
  \label{tab:numerical}}

\section{High-Energy Behavior and Unitarity}

The form of the elastic pion-pion scattering amplitudes at high energies is particularly 
interesting because it can potentially yield an upper bound on the scale at which the 
effective field theory becomes strongly coupled.  
The two-body  amplitude for elastic pion-pion scattering
lies on the ``Argand" circle of radius $1/2$ centered, in the complex plane, at the point $i/2$.   In the Born (tree-level) approximation, 
however, the amplitude is {\it always} real.   As the Born
level amplitude deviates further from the Argand circle,  higher-order corrections
must become larger (in absolute value) in order for the full amplitude to respect unitarity.  
Therefore, the extent to which the tree-level amplitude departs from the Argand circle 
may be used as a measure of how strongly-coupled the theory has become. A common choice for the criterion by which a theory is judged to have become strongly coupled is
\begin{equation}
T^{born} \simeq {\rm Re}(T) \ge \frac{1}{2}~.
\label{eq:bornunitarity}
\end{equation}
The rationale for this choice 
is that in order for the full amplitude
to lie on the Argand circle,  higher order corrections
must have a magnitude of order 1/2 as well, making the tree-level and loop-level
contributions comparable.

We have seen, surprisingly, that the 
behavior of the high-energy amplitudes is drastically changed by including the exchange of the $KK$ modes. In general terms, these results are consistent with those found in QCD 
\cite{Sannino:1995ik} and in previous investigations of Higgsless models 
\cite{Papucci:2004ip}.  In this section we plot the leading partial-wave scattering amplitudes in a flat space model for several values of the lightest $KK$ mode's mass and discuss the implications.

\subsection{s-wave Isosinglet Elastic Pion Scattering}

\DOUBLEFIGURE[th]{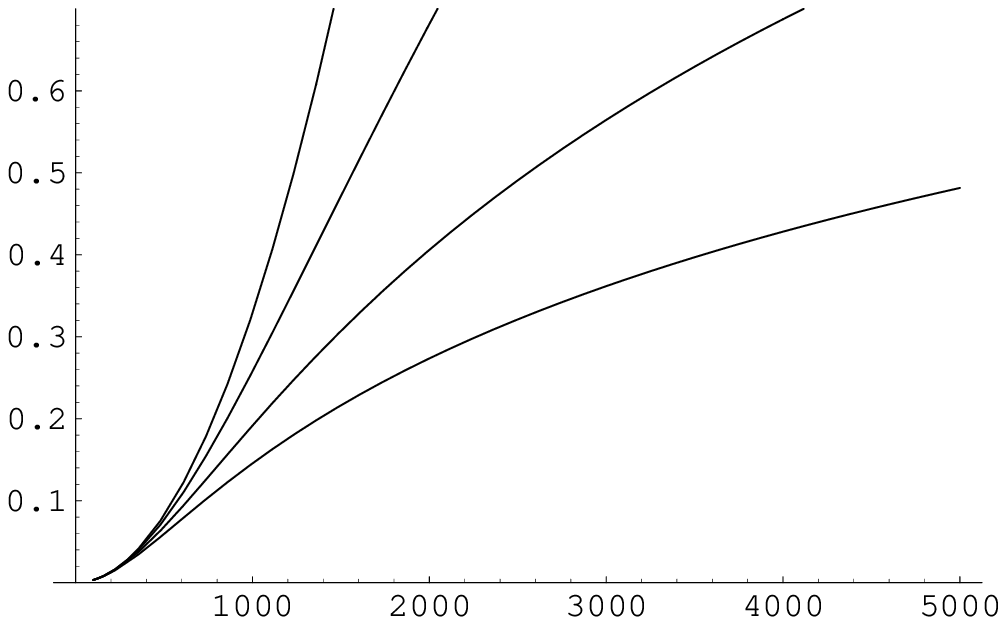,width=0.45\textwidth}{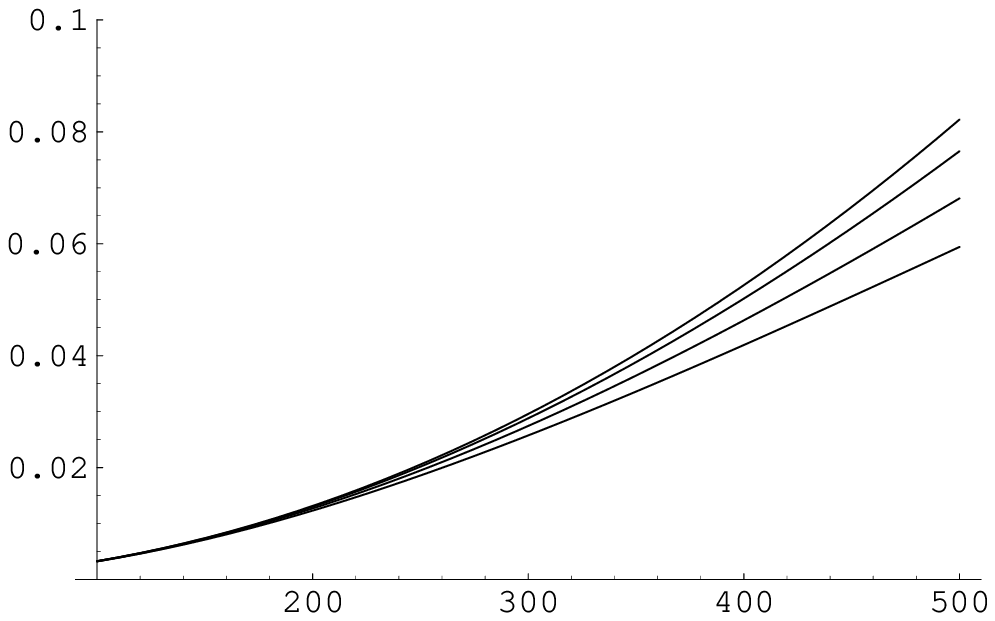,width=0.45\textwidth}
{Plot of the amplitude $T^0_0(E)$ for -- from lowest to highest --
$M_{W1}= 500, 700, 1200$  GeV and $M_{W1}=\infty$
(the low-energy theorem). The units on the horizontal axis are GeV. \label{fig:tthree}}
{Expanded view of the low-energy portion of Fig. 6; note that all amplitudes show the correct low-energy behavior.}

We expect that the largest partial wave for elastic pion-pion scattering will correspond to spin-0, isospin-0 scattering; the analytic expression for this partial wave amplitude in a flat-space continuum $SU(2)$ model is given in Eq. (\ref{eq:T00}). The values of $T^0_0(s)$ corresponding to $M_{W1} = 500,700, 1200$ GeV and $M_{W1}=\infty$ are plotted in Fig. \ref{fig:tthree}, and are similar to those in \cite{Foadi:2003xa}. Fig. 7 shows an expanded view of the low-energy behavior of the scattering amplitudes in Fig. \ref{fig:tthree}; we can see by inspection that all curves have the same low-energy behavior, as expected. Conversely, the asymptotic behavior of the amplitude at high energies shows logarithmic growth with  $s$, as in Eq. (\ref{eq:asymp00}).

\begin{figure}[hbt]
 \label{fig:Fig7}
 \begin{center}
 \includegraphics[width=7.5cm,clip=true
                 ]{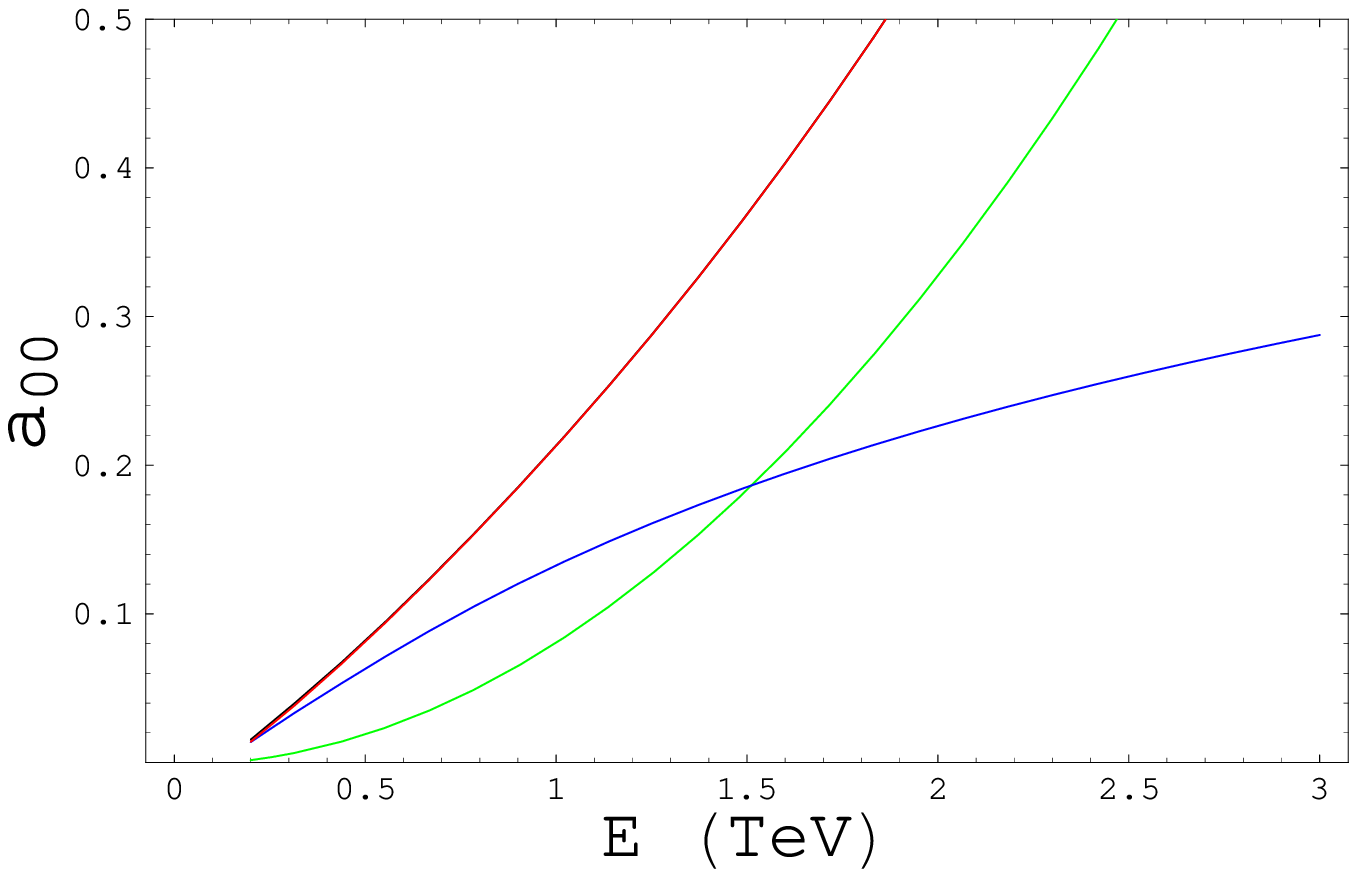}\ 
 \includegraphics[width=7.5cm,clip=true
                 ]{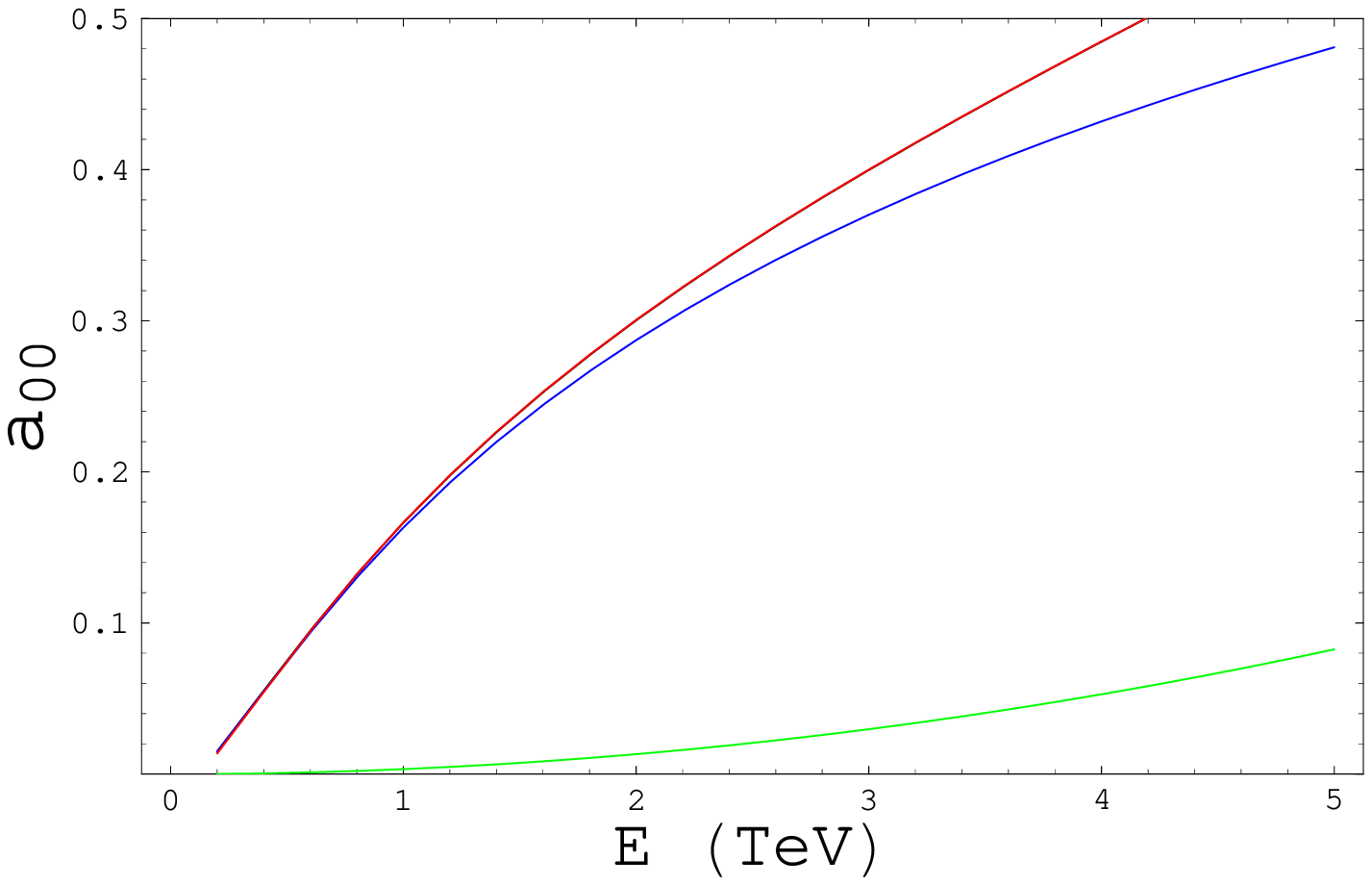}
 \vspace*{-1mm}
 \caption{Unitarity limits for elastic scattering:
 $\,W_{0L}W_{0L}\to W_{0L}W_{0L}$ and 
   ${{\pi}}_0 {{\pi}}_0\to {{\pi}}_0 {{\pi}}_0\,$.\,
 The $s$-wave amplitude
 $a_{00}^{~}$ in the $I=0$ channel is plotted as a function of c.m.\
 energy $E$.\, The left (right) plot is for
 $N\!+\!1=2$ ($N\!+\!1=10$) Higgsless
 deconstruction with charged-gauge-boson masses of 
 $(M_{W0},\,M_{W1})=(80,\,500)$\,GeV.
 In each plot, the top curve (red) is the full
 longitudinal gauge boson amplitude which almost exactly overlaps with
 the corresponding full Goldstone boson amplitude (black curve),
 confirming the Equivalence Theorem for elastic zero-mode scattering.
 The middle curve (blue) is the Goldstone boson amplitude
 including the $(t,u)$-channels only, while the
 bottom curve (green) is the Goldstone boson amplitude
for the four-pion contact diagram only.
 }
 \end{center}
 \setlength{\unitlength}{1mm}
 \begin{picture}(0,0)
 \thicklines
 \put(15,90){$N\!+\!1=2$}
 \put(50,68){$t/u$-Channels}
 \put(34,57){Contact}
 \put(32,82){Full}
 \put(90,90){$N\!+\!1=10$}
 \put(110,72){$t/u$-Channels}
 \put(110,57){Contact}
 \put(110,82){Full}
 \end{picture}
 \label{fig:newsix}
 \end{figure}

By way of comparison, Fig. 8 shows the amplitudes for s-wave isosinglet $\,W_{0L}W_{0L}\to W_{0L}W_{0L}$ and 
   ${{\pi}}_0 {{\pi}}_0\to {{\pi}}_0 {{\pi}}_0\,$ scattering in a 3-site and an 11-site deconstructed model, each with $M_{W1}=500$ GeV. Note that one should compare the continuum calculation of Fig. 6 to
the ``$t/u$-Channels" contributions plotted in Fig. 8 since the contact interaction
contribution of the deconstructed models vanishes in the continuum limit.

From Fig. \ref{fig:tthree} we may immediately read off the elastic unitarity limit (energy scale where $T^0_0 = \frac{1}{2}$) for the flat-space model.  
For $M_{W1} = 500$ (700, 1200) GeV, we obtain an elastic unitarity limit 
 of order 5.4 (2.6, 1.6) TeV.

\subsection{Significance of Unitarity Bounds}

 The asymptotic form of the elastic s-wave isosinglet scattering amplitude shown in Eq. (\ref{eq:asymp00}) and in figs. 6-8 is quite sensible in one context: the
pion scattering amplitude arises ultimately from the 5-D gauge interaction, whose strength
is given by $\tilde{g}$.  The logarithmic
growth of the amplitude is a reflection of the $t$-channel pole that would have been present
in the amplitude, were there any massless particles. The form of Eq. (\ref{eq:asymp00}) also explains  why one finds a scale of unitarity
violation that grows so dramatically as the mass $M_{W1}$ is reduced:
 the coefficient of the logarithm becomes smaller as ${\mathsf m}_{W1}$ is lowered 
and therefore the scale $s$ at which the amplitude ${\cal T}^0_0(s)$ violates
unitarity grows exponentially.

A separate question is the physical meaning of the exponentially growing scale at which the $s$-wave
tree-level amplitudes violate unitarity.  For models in which the lightest $KK$ mass is below about 700 GeV, the scale of elastic unitarity violation is clearly above the threshold for $\pi\pi \to W_1 W_1$ inelastic scattering.  Moreover, due to phase space considerations, we expect that the inelastic channels will dominate when available.  Hence a more accurate assessment of the scale at which two-body scattering violates unitarity must be performed in the context of a coupled-channels analysis including inelastic channels (see e.g. \cite{Papucci:2004ip}).  We will report on such an analysis in a forthcoming work \cite{ref:futureref}.

\section{Conclusions}

We have applied the Kadanoff-Wilson block-spin transformation to a deconstructed 5-dimensional gauge theory and showed how this enables us to obtain an alternative deconstruction with fewer factor groups, yet exhibiting the same low-energy properties.  Moreover, we found that the freedom to perform the block-spin transformation corresponds, in the continuum limit, to the freedom to describe the continuum theory in different coordinate systems.  In particular, we demonstrated how to perform an f-flat deconstruction in which all of the f-constants of the linear moose are identical.

We then applied these findings to enable us to study the properties of $W_L W_L$ elastic scattering in $SU(2)^2 \times U(1)$ Higgsless models.   If one performs an f-flat deconstruction of these models, the phenomenologically relevant limit is that in which the gauge couplings of the end sites ($g_0, g_{N+1}$) are small \cite{Georgi:2004iy,SekharChivukula:2004mu}.  Moreover, the equivalence theorem tells us that (at energies far above the $W$ boson mass) scattering of longitudinally polarized electroweak gauge bosons in the case of small $g_0,\ g_{N+1}$ corresponds to the scattering of the pions in the extreme limit $g_0,\, g_{N+1} \to 0$.  Accordingly we have studied elastic pion-pion scattering in the global linear moose as a way of understanding $W_L W_L$ elastic scattering in Higgsless models with arbitrary background 5-D geometry, spatially dependent gauge-couplings, and brane kinetic energy terms for the gauge-bosons. 

We have derived the form of the general amplitudes and the leading partial-wave amplitudes for elastic pion-pion scattering in global linear moose models and examined their limiting forms in the case of scattering energies well below or well above the mass of the lightest KK modes.  We applied these results directly to continuum $SU(2)$ and $SU(2)\times SU(2)$ models in both flat and warped space.  We also confirmed an alternative formulation of the low-energy scattering amplitudes in terms of the Longhitano parameters
 
 The form of the elastic pion-pion scattering amplitudes at high energies is particularly 
interesting because it can potentially yield an upper bound on the scale at which the 
effective field theory becomes strongly coupled.  We have studied the behavior of the largest partial wave amplitude (spin-0, isospin-0 scattering) in a flat-space continuum $SU(2)$ model and its three-site and eleven-site f-flat deconstructions.  We conclude that elastic unitarity provides a useful guide to the range of a model's validity only for models in which the lightest KK mode has a mass greater than about 700 GeV; otherwise, inelastic channels such as  $\pi\pi \to W_1 W_1$ become available and important.  In models where $M_1 \leq 700$GeV, a more accurate assessment of the scale at which two-body scattering violates unitarity requires a coupled-channels analysis including inelastic channels.

\centerline{\bf Acknowledgments}

R.S.C. and E.H.S. are supported in part by the US National Science Foundation under
grant  PHY-0354226.  M.T.'s work is supported in part by the JSPS Grant-in-Aid for Scientific Research No.16540226. H.J.H. is supported by Tsinghua University.  M.K. is supported in part by the US National Science Foundation under grant PHY-0354776.  R.S.C. and E.H.S. thank the Aspen Center for
Physics for its hospitality while this work was being completed.


\end{document}